# Activity stabilization in a population model of working memory by sinusoidal and noisy inputs

**Nikita Novikov[1*], Denis Zakharov[1], Victoria Moiseeva[1], Boris Gutkin[1,2]**

[1]Centre for Cognition and Decision Making, HSE University, Moscow, Russia

[2]Group for Neural Theory, LNC2 INSERM U960, Départment d'études cognitives, École Normale Supérieure, PSL Research Université, Paris, France

*** Correspondence:**
nikknovikov@gmail.com, n.novikov@hse.ru

## Abstract

According to mechanistic theories of working memory (WM), information is retained as stimulus-dependent persistent spiking activity of cortical neural networks. Yet, how this activity is related to changes in the oscillatory profile observed during WM tasks remains a largely open issue. We explore joint effects of input gamma-band oscillations and noise on the dynamics of several firing rate models of WM. The considered models have a metastable active regime, i.e. they demonstrate long-lasting transient post-stimulus firing rate elevation. We start from a single excitatory-inhibitory circuit and demonstrate that either gamma-band or noise input could stabilize the active regime, thus supporting WM retention. We then consider a system of two circuits with excitatory intercoupling. We find that fast coupling allows for better stabilization by common noise compared to independent noise and stronger amplification of this effect by in-phase gamma inputs compared to anti-phase inputs. Finally, we consider a multi-circuit system comprised of two clusters, each containing a group of circuits receiving a common noise input and a group of circuits receiving independent noise. Each cluster is associated with its own local gamma generator, so all its circuits receive gamma-band input in the same phase. We find that gamma-band input differentially stabilizes the activity of the "common-noise" groups compared to the "independent-noise" groups. If the inter-cluster connections are fast, this effect is more pronounced when the gamma-band input is delivered to the clusters in the same phase rather than in the anti-phase. Assuming that the common noise comes from a large-scale distributed WM representation, our results demonstrate that local gamma oscillations can stabilize the activity of the corresponding parts of this representation, with stronger effect for fast long-range connections and synchronized gamma oscillations.

## 1 Introduction

The concept of working memory (WM) characterizes the ability of the brain to retain in an active form certain information that is relevant to a current task but is not perceived at the particular moment by sensory systems (Baddeley, 2003). One of the main mechanisms supposedly underlying WM is self-sustained activity of neural populations (Goldman-Rakic, 1995; Compte, 2006), which can be turned on or off in a short period of time (about 100 ms) and continue for a time interval of up to tens of seconds (Wang, 2001). Cells whose activity is maintained at an elevated level during the retention of information in WM have been found in various parts of the brain, primarily in the prefrontal cortex (Fuster, Alexander, 1971; Funahashi et al., 1989; Miller et al., 1996; Chafee, Goldman-Rakic, 1998; Constantinidis, Goldman-Rakic, 2002).

The process of retaining information in WM is also associated with changes in the collective rhythmic activity of brain networks (which are neuronal oscillations) in various frequency bands (Sauseng et al., 2009; Siegel et al., 2009; Haegens et al., 2010; Liebe et al., 2012; Wimmer et al., 2016; Lundqvist et al., 2016; Lundqvist et al., 2018; Kornblith et al., 2016). Among these changes, increase of the gamma-band activity is of special interest. Gamma activity usually reflects activation of neural populations and coincides with episodes of firing rate elevation. In WM tasks, gamma activity increases most strongly during presentation of stimuli (Wimmer et al., 2016; Lundqvist et al., 2016; Kornblith et al., 2016), but in the delay period (i.e. in the time period after stimulus termination and before an instruction to make a response) it is still higher than in the baseline (Jokisch, Jensen, 2007; Palva et al., 2011; Lutzenberger et al., 2002; Kaiser et al., 2003; Haegens et al., 2010; Wimmer et al., 2016; Lundqvist et al., 2016; Kornblith et al., 2016). We note that during this delay period, the subject has to withhold the response, yet needs to retain "on-line" information necessary to generate the appropriate response. The delay-period gamma activity presumably reflects activation of the neural populations that represent the WM content (Roux, Uhlhaas, 2014). This is supported by the findings that the delay-period gamma activity is higher than in the passive observation task (Wimmer et al., 2016), increases with WM load (Howard et al., 2003; van Vugt et al., 2010; Kornblith et al., 2016; Lundqvist et al., 2016) and only in the task-relevant regions (Kaiser et al., 2003; Jokisch, Jensen, 2007) or at those cortical sites that contain neurons selective to the WM content (Kornblith et al., 2016; Lundqvist et al., 2016).

Recently, the delay-period gamma oscillations were more directly linked to activation of WM representations. It was demonstrated that gamma activity is irregular at the single-trial level, and the episodes of increased gamma power ("gamma-bursts") are associated with elevated firing rates and increased amount of information about the WM content that could be decoded from spiking activity (Lundqvist et al., 2016; Lundqvist et al., 2018; Bastos et al., 2018). It is not fully clear, however, whether the gamma power increase plays a functional role in WM retention or is it merely a consequence of transient firing rate increase during spontaneous reactivations of WM representations.

Besides the gamma power increase, an increase in gamma-band coherence between different cortical sites during the delay period was reported (Lutzenberger et al., 2002; Kaiser et al., 2003; Palva et al. 2010; Kornblith et al., 2016). Furthermore, it was shown that transcranial gamma-band electrical stimulation of two distant sites could improve performance in a WM task (Tseng et al., 2016). Interestingly, the improvement was observed only under anti-phase (but not under in-phase) stimulation. This result suggests that gamma-band coherence presumably plays a functional role in WM retention, and is not merely an epiphenomenon.

Nowadays, a number of computational WM models exist. Most of them are based on multistable neural networks. In the simplest case, a system has two stable states, one of which (with low firing rates) corresponds to the background regime, and the other one (with higher firing rates) relates to the active regime, in which an object is retained in WM. Transition from the background to the active state occurs under the action of a short excitatory pulse that mimics the arrival of a to-be-memorized stimulus. As an alternative, there are models, in which the active retention regime is metastable, and the system slowly returns to the background state after a stimulus presentation (Lim, Goldman, 2013). In many WM models, the self-sustained post-stimulus firing rate elevation is provided by reverberation of excitation in the network due to synaptic interactions (Amit, Brunel, 1997; Brunel, Wang, 2001). In addition, there are models in which post-stimulus enhancement of synaptic connections due to short-term plasticity plays a significant role (Mongillo et al., 2008; Mongillo et al., 2012; Hansel, Mato, 2013).

Many theoretical papers, following Amit and Brunel (1997), described WM models with asynchronous spiking activity. It was shown that if a model contains only fast excitation, then even slight synchronization returns it to the background state (Laing and Chow, 2001; Gutkin et al., 2001; Compte, 2006). However, generation of oscillations (i.e. synchronization) in a WM model is possible in the presence of slow NMDA receptors (Tegner et al., 2002) or in the case of modular structure of the system (Lundqvist et al., 2010; Lundqvist et al., 2011). Besides investigating the mechanisms of oscillations' appearance in WM models (Tegner et al., 2002; Roxin and Compte, 2016), several theoretical studies explore possible functional roles of oscillations in WM (Lisman Idiart, 1995; Ardid et al., 2010; Lundqvist et al., 2010; Kopell et al., 2011; Lundqvist et al., 2011; Chik, 2013; Dipoppa, Gutkin, 2013; Pina et al., 2018; Schmidt et al., 2018; Sherfey et al., 2020). The methodology, however, differs substantially between these studies, and a unified theoretical framework in this field is still lacking.

We follow a paradigm used in (Dipoppa and Gutkin, 2013; Schmidt et al., 2018), in which a WM system is resonant and receives an external oscillatory input that controls its behavior. Schmidt et al. (2018) showed that gamma-band input could stabilize WM retention even if the active regime is initially metastable (which means that the firing rate increases after stimulus presentation, but then slowly returns to the background level in the absence of input oscillations). In the present study, we make a step further and consider systems with metastable active regime that are comprised of several excitatory-inhibitory circuits coupled by symmetrical excitatory connections and receiving the same stimulus-related signal (such systems could serve as models of a distributed representation of an object in WM). We explore stabilization of the metastable active regime by external gamma-band and white-noise inputs, both of which are assumed to come from neural populations not explicitly included into the model.

We assume that the gamma-band inputs are generated locally (and thus could have different phases for different circuits), and that the noise inputs originate from distributed networks (that could jointly participate in WM retention, thus producing correlated activity). The model parameters in the focus of our research are the following: (1) in-phase or anti-phase character of the gamma inputs to the circuits, (2) commonality or independence of the noise inputs to the circuits, (3) NMDA:AMPA ratio of inter-circuit connections. By varying these parameters, we aim to understand whether gamma-band oscillations are able to preferentially stabilize the active regime in those circuits that participate in collective WM-related activity (and thus receive a common noise input rather than independent inputs). Another goal of our study is to explore whether synchronization of gamma generators (leading to in-phase gamma inputs to different circuits) affects the ability of the gamma input to stabilize WM retention, depending on whether the inter-circuit connection in the system are fast or slow.

The paper is organized as follows. We start from a single-circuit model and demonstrate that the active regime could be stabilized by gamma-band or white-noise input. Next, we explore joint stabilizing effect of gamma-band and white-noise inputs in a system of two circuits with mutual excitation. We vary the NMDA:AMPA ratio of the inter-circuit connections and parameters of the inputs (including phase difference between the gamma inputs to the circuits and commonality / independence of the noise inputs); for each parameter combination we evaluate the effectiveness of stabilization. Finally, we consider a multi-circuit system comprised of two local clusters, with all the circuits in a cluster receiving the same gamma input. Each cluster, in turn, contains two circuit groups: the circuits from the first groups receive a common noise input and the circuits from the second groups – independent noise inputs. We explored the stabilizing effect of gamma input on each circuit group, depending on whether it is delivered to the clusters in the same phase or in the anti-phase, conditioned on the type (slow / fast) of the inter-cluster connections.

## 2 MATERIALS AND METHODS

### 2.1 Model description

In this paper, we consider firing rate models of WM that consist of one, two, or many circuits each containing interacting excitatory and inhibitory neural populations (Fig. 1). These circuits serve as representations of various parts or features of WM content and are linked by symmetrical excitatory connections. Each circuit receives an external input consisting of several components: (1) tonic (constant) input, (2) stimulus-related input, (3) zero-mean sinusoidal oscillatory input, and (4) white noise. The stimulus-related input is implemented as a rectangular pulse whose amplitude and duration are the same for all circuits in the model. The stimulus input is projected 80% to the excitatory population of a circuit and 20% to the inhibitory population. The oscillatory input impinges only on the excitatory populations of circuits and represents a modulatory signal from the outside of the modelled WM network. White noise mimics the input produced by activity of a larger network into which our model is embedded, but which was not modelled explicitly. We explore different cases, in which the circuits receive in-phase or anti-phase oscillatory inputs, as well as identical or independent noisy inputs.

The state of circuit populations is described by the following dynamical variables: (1) firing rates, (2) mean input currents (AMPA, NMDA, and GABAA), and (3) population variances of the input currents (AMPA and GABAA). We further argue that if the mean values of the AMPA and NMDA currents are of the same order of magnitude, the NMDA current variance has to be much smaller than the AMPA variance, since the NMDA time constant is much larger than the AMPA time constant. Hence, we set the NMDA current variance to zero. Thus, our model of interacting circuits is described by the following equations:

$$\begin{cases} \tau_{ra} \dfrac{dr_a^p}{dt} = -r_a^p + F_{ra}(\mu_a^p, \sigma_{a,AMPA}^p, \sigma_{a,GABAA}^p) \\ \mu_a^p = \sum\limits_S \mu_{a,S}^p, \quad S = AMPA, NMDA, GABAA \\ \tau_S \cdot \dfrac{d\mu_{a,S}^p}{dt} = -\mu_{a,S}^p + \tilde{\mu}_{a,S}^{p,rec} + \tilde{\mu}_{a,S}^{p,cc} + \tilde{\mu}_{a,S}^{p,ext} \\ \dfrac{\tau_S}{2} \cdot \dfrac{d\left(\sigma_{a,S}^p\right)^2}{dt} = -\left(\sigma_{a,S}^p\right)^2 + \left(\tilde{\sigma}_{a,S}^{p,rec}\right)^2 + \left(\tilde{\sigma}_{a,S}^{p,cc}\right)^2 + \left(\tilde{\sigma}_{a,S}^{p,ext}\right)^2 \end{cases},$$

where the lower index $a$ denotes a population type ($e$ – excitatory, $i$ – inhibitory), the upper index $p$ denotes a circuit number. The variable $r_a^p$ is the firing rate; $\mu_a^p$ is the total mean input current, $\mu_{a,S}^p$ is the mean input current via the synapses of the type S (S denotes AMPA, NMDA, or GABAA); $\left(\sigma_{a,S}^p\right)^2$ is a population variance of the input current via the synapses of the type S; $\tau_{ra}$ is the time constant that governs the firing rate dynamics, $\tau_S$ is the synaptic time constant for the synapses of the type S; $\tilde{\mu}_{a,S}^{p,rec}$ , $\left(\tilde{\sigma}_{a,S}^{p,rec}\right)^2$ are the population mean and variance of the recurrent (intra-circuit) inputs; $\tilde{\mu}_{a,S}^{p,cc}$ , $\left(\tilde{\sigma}_{a,S}^{p,cc}\right)^2$ are the population mean and variance of the input from other modelled circuits; $\tilde{\mu}_{a,S}^{p,ext}$ and $\left(\tilde{\sigma}_{a,S}^{p,ext}\right)^2$ are the population mean and variance of the external inputs; $F_{ra}$ are the gain (transfer) functions. Note that the time constant for the population variance of each currents is twice smaller than the time constant for the population mean of the same current (due to the properties of linear first-order stochastic ODE's; see (Renart et al., 2007) for an example of a model with dynamical mean and variance of the input current). Within-circuit excitatory-to-excitatory connections demonstrate short-term plasticity (Tsodyks, Markram, 1997), see the full description in the supplementary materials.

We also need to define the gain functions for the neural populations. Instead of defining these gain functions $F_{re}$ and $F_{ri}$ apriori, we calculate them following an approach similar to the one previously used in (Schaffer et al., 2013; Augustin et al., 2017) that allows to match low-dimensional models with spiking networks, making them more biologically plausible. According to this approach, the gain functions were pre-calculated on a cubic grid (with the coordinates: total mean current, AMPA current std., GABAA current std.) by numerical simulations of leaky integrate-and-fire (LIF) neurons having typical properties of the regular-spiking (pyramidal) neurons and the fast-spiking interneurons of the cortex. For an arbitrary point, the value of a gain function was obtained by interpolation between the pre-calculated values at the closest grid nodes. To show the shape of the gain functions, in Fig. 2C,D we present their projections to the plane of mean firing rate and total mean input current. The projections were made for two different combinations of AMPA and GABAA current std. that correspond to the background and active states of the bistable single-circuit system (these states are depicted in Fig. 2A; see the next section for details). Since our model is tuned to operate in a subthreshold regime (which is typical for cortical neurons; (Compte et al., 2003; Wang, 2010)), the aforementioned plots have an exponential-like, concave shape.

The full system of model equations and its detailed description are given in the Supplementary Materials.

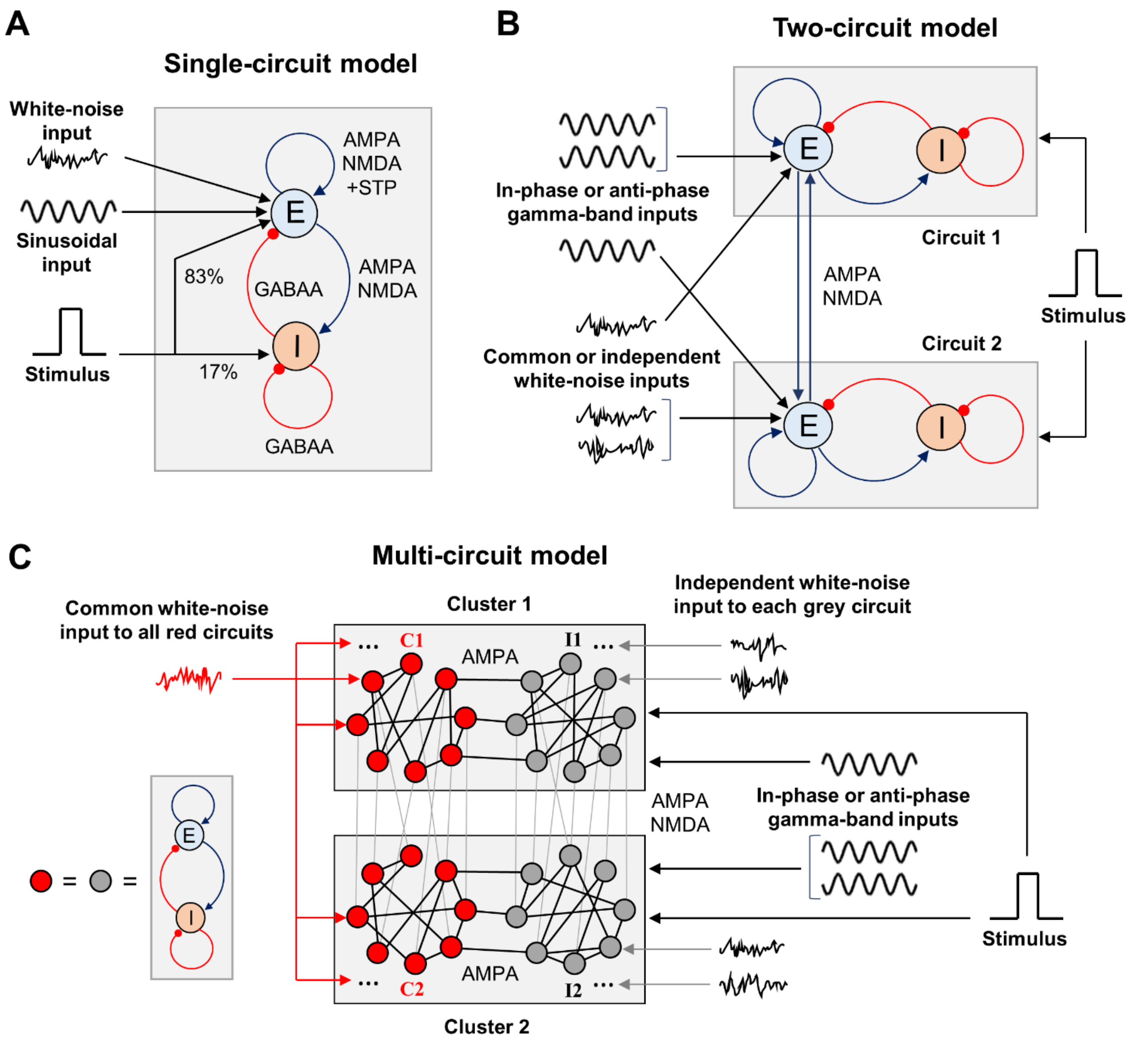

Figure 1. Schematic representation of the WM model types we explore in this study. (A) Single-circuit system. E – excitatory population, I – inhibitory population. The excitation is mediated by AMPA and NMDA receptors, inhibition – by GABAA receptors. The excitatory-to-excitatory connections demonstrate short-term plasticity. White-noise and oscillatory inputs are delivered to the excitatory population. A rectangular stimulus-related signal is delivered to the excitatory and inhibitory populations in the proportion of 5:1. Both populations also receive constant inputs (not shown). (B) Two-circuit system. The circuits are structurally identical to the one shown in (A) and intercoupled via excitatory connections with varying NMDA:AMPA ratio. Gamma-band sinusoidal input is delivered to the excitatory populations of the circuits, either in the same phase or in the anti-phase. White-noise inputs (identical or independently generated) are also delivered to the excitatory populations. The stimulus-related signal is delivered in the same way as in (A), and it is identical for both circuits. (C) Multi-circuit model. Each circle represents a circuit, structurally identical to the one shown in (A). The circuits are grouped into two local clusters; each cluster contains two circuit groups. Lines connecting the circles represent symmetrical excitatory inter-circuit connections. Within-cluster inter-circuit connections are mainly AMPA-mediated (NMDA:AMPA ratio is 0.1), inter-cluster connections could be predominantly mediated either by AMPA or NMDA. The circuits from the “red” groups (C1, C2) receive a common white-noise input; the circuits from the “grey” groups (I1, I2) receive independent white-noise inputs.

All the circuits from each cluster receive gamma-band input in the same phase; the inputs to the clusters could have the same phase or the opposite phases. The same stimulus-related signal is delivered to each circuit.

## 2.2 Model parameters

*Single circuit*

We set the parameters of a circuit (Fig. 1A) in such way that: (1) it has a stable background steady-state with a low level of activity, (2) it responds to a stimulus by prolonged activity increase with subsequent return to the vicinity of the background state, and (3) it has gamma-band resonance during the post-stimulus increased activity. We refer to the increased post-stimulus activity as metastable active regime and refer to a system with such regime as metastable system.

To obtain the required behavior of the circuit (metastability and gamma-band resonance), we set the parameters in the following way. We start by pre-selecting parameters that provide bistability in a circuit. In this case, there are three equilibria in the phase space: two stable ones (corresponding to the background and active states) and a saddle (Fig. 2B). We tuned the parameters in such way that the second (active) equilibrium has a pair of complex-conjugate eigenvalues with a small negative real part and an imaginary part corresponding to the gamma band. An orbit of such a system, when starting near the active steady state, shows slowly decaying gamma-band oscillations returning back to the active state (an example is presented in Fig. S1B). Then, we decreased the weight of the excitatory-to-excitatory synaptic connections, until the upper equilibrium disappears through a fold bifurcation (see the phase plane with the single steady state in Fig. 2A). In fact, the upper equilibrium leaves a “ghost” near which the dynamics are slow. Thus, the resulting system is metastable, and the slow dynamics near the “ghost” corresponds to the metastable active regime. This regime inherits gamma-band resonance from the active steady state that existed in the bistable system before the bifurcation. An example orbit of the metastable system showing damped oscillations with subsequent decay to the background is presented in Fig. S1A.

Finally, we explored the ability of white-noise and sinusoidal inputs to stabilize the metastable active regime.

*Two circuits*

We also considered a system of two identical interacting circuits (Fig. 1B). In this case, we symmetrically connect the circuits using excitatory coupling between their excitatory populations. At the same time, to compensate this additional inter-circuit excitation, we decreased the background inputs to the excitatory populations. In this way, we keep the metastable behavior in each circuit and could study the influence of input oscillations and noise.

We varied NMDA:AMPA ratio for the inter-circuit connections and explored the ability of in-phase / anti-phase oscillatory inputs and common / independent white-noise inputs to stabilize the active regime. We used duration of increased post-stimulus activity as the measure of the active regime stability. The post-stimulus activity of a circuit was considered to be terminated when the time-course of its excitatory population firing rate smoothed with the 100 ms time window fell below the level of 3 Hz.

*Multiple circuits*

Finally, we considered a multi-circuit system schematically presented in Fig. 1C. Each circuit is represented by a (red or black) circle. Links between the circles correspond to mutual excitatory connections between the circuits. In the system considered, there were two circuit clusters that mimic two spatially separated local networks. All circuits in each cluster receive input oscillations of the same phase, whereas the circuits in different clusters may receive either in-phase or antiphase oscillations.

Each cluster contains two groups of circuits. Circuits from the first groups C1, C2 (red-colored circuits in both clusters in Fig. 1C) receive a common noise input. Circuits from the second groups I1, I2 (gray circuits in Fig. 1C) receive independent noise inputs.

Each group in our model is a random graph with eight nodes (circuits) having the constant in-degree of three; two groups within the same cluster (C1 – I1, C2 – I2) are connected by three randomly assigned links; corresponding groups in different clusters (C1 – C2, I1 – I2) are connected by eight randomly assigned links (with one link per each node); non-corresponding groups in different clusters (C1 – I2, C2 – I1) are not connected. All links between circuits are bi-directional.

We considered two multi-circuit models: one with the fast inter-cluster connections (90% AMPA, 10% NMDA), and the other one with slow inter-cluster connections (100% NMDA). All other parameters were identical. We simulated both models in several regimes: (1) no oscillatory input, (2) only one cluster receives oscillatory input, (3) the clusters receive in-phase oscillations, (4) the clusters receive anti-phase oscillations. For each regime, we were interested in the average duration of post-stimulus activity of each circuit group (C1, I1, C2, I2).

## 3 Results

### 3.1 Activity of single-circuit model

A single-circuit model consisting of an excitatory and an inhibitory population (Fig. 1A) could be either bistable or monostable, depending on the strength of the self-excitation. For strong enough self-excitation, the model has three equilibria (Fig. 2B): two stable steady states and a saddle. The stable equilibria correspond to two types of activity: the background (low-rate) and active (high-rate) regimes, while the saddle manifold separates their basins of attraction. For weaker self-excitation, the equilibria that correspond to the active state and the saddle disappear through the fold bifurcation, and a metastable active regime (with high firing rate) remains as a ghost of the active state (Fig. 2A). In this case, the circuit could be excited from the background state by a stimulus and demonstrate relatively long transient high-firing-rate activity.

The metastable active regime can be stabilized (i.e. the decay of the post-stimulus activity could be slowed or prevented) either by an oscillatory signal (Fig. 3), or by a noisy input (Fig. 4). In the case of oscillatory input, there is a range of its amplitudes, for which the stabilization occurs only if the input frequency falls into the lower gamma band (middle part of Fig. 3A). For lower amplitudes, the stabilization does not occur (grey region in Fig. 3A). For higher amplitudes, the frequency range of stabilization expands to the high gamma and beta bands. If the input amplitude is too high (above the dashed red line in Fig. 3A), the background regime disappears, i.e. the oscillations put the system into the active (high-firing-rate) regime even without stimulus presentation.

We found that noisy input is also able to stabilize the metastable state (Fig. 4). The purple curve represents the mean firing rate observed in the absence of a stimulus, as a function of the noise standard deviation. The green curve represents the mean post-stimulus firing rate. At low intensity, the noise was unable to stabilize the active regime, while at high intensity, the noise put the system to the active regime even without stimulus presentation. Thus, an intermediate range of noise intensities is appropriate for WM functioning; it is seen in Fig. 4A as the range in which the green line goes considerably above the purple line. For a value from this range, the persistent activity is initiated by a transient stimulus, and the WM trace is then kept "alive" by the noise input (Fig 4B). At the same time, the system stays in the background regime if the stimulus is not presented (Fig. 4C).

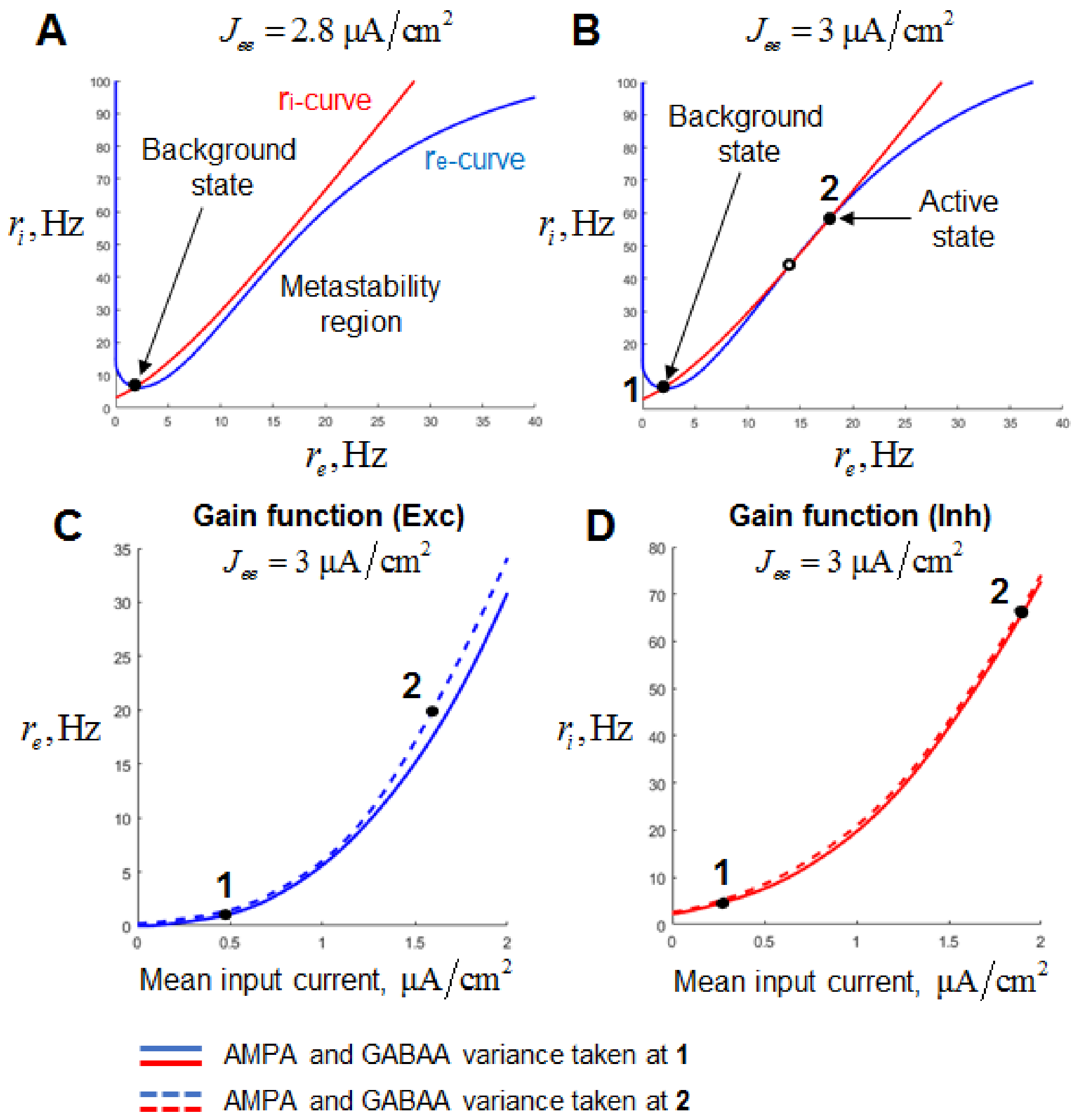

Figure 2. Steady-states of a single-circuit model. (A) Phase plane for the single-circuit model analyzed in our study. For the points on the blue curve (re-curve), all the derivatives, except of dri/dt are zero; for the point on the red curve (ri-curve), all the derivatives, except of dre/dt are zero. The system has the single (background) steady-state and the region of slowly decaying activity (metastability region), where the re- and ri-curves are close to each other. (B) Phase plane for a system with increased recurrent excitation. The system has the background steady-state (denoted as 1) and the active steady-state (denoted as 2). (C) Two slices of the gain function of the excitatory population, representing dependence of the excitatory firing rate on the mean input current to the excitatory population. Solid curve – the slice taken at the constant values of the AMPA- and GABAA-current variance, equal to the values calculated at the steady-state 1. Dashed curve – the slice taken at the variance values calculated at the steady-state 2.
(D) Same as (C), but for the inhibitory population.

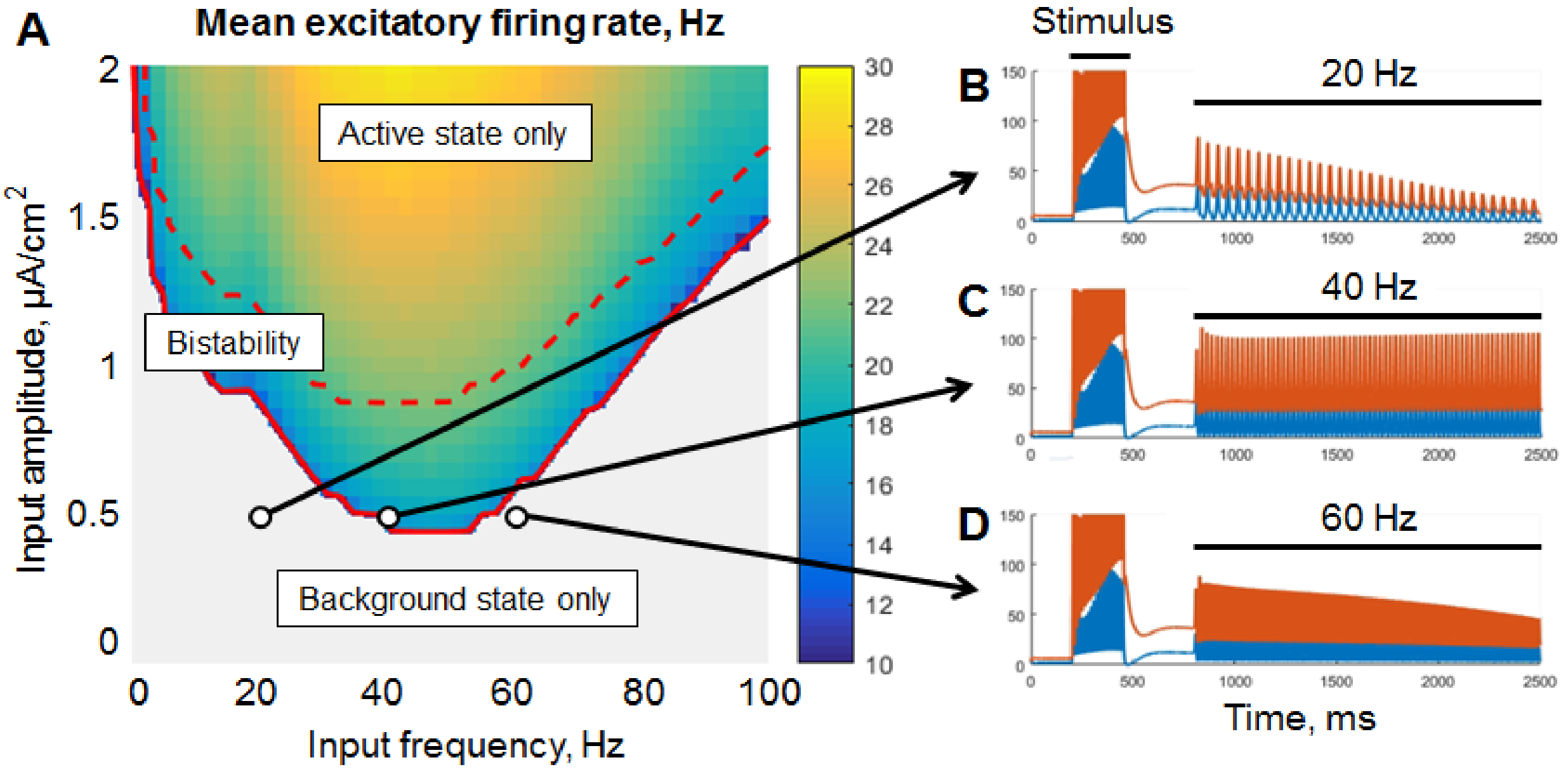

Figure 3. Resonant stabilization of the metastable activity by a sinusoidal signal (without noise) in the single-circuit model. In panel (A), there is a diagram showing the mean excitatory firing rate for different values of the amplitude and frequency of the signal. Simulation was performed for 10 seconds, and the firing rate was averaged over the last period of the input signal. Above the solid red curve, the active (initially metastable) regime is persistent; above the dashed red curve, there is no background regime. (B,C,D) –examples of the activity traces. Blue – excitatory firing rate, red – inhibitory firing rate. The system was shifted to the metastable active regime by a transient constant input (stimulus), after which an oscillatory input was turned on (20, 40, or 60 Hz zero-mean sinusoid). Note that the 40 Hz periodic input (panel C) provided a persistent active regime

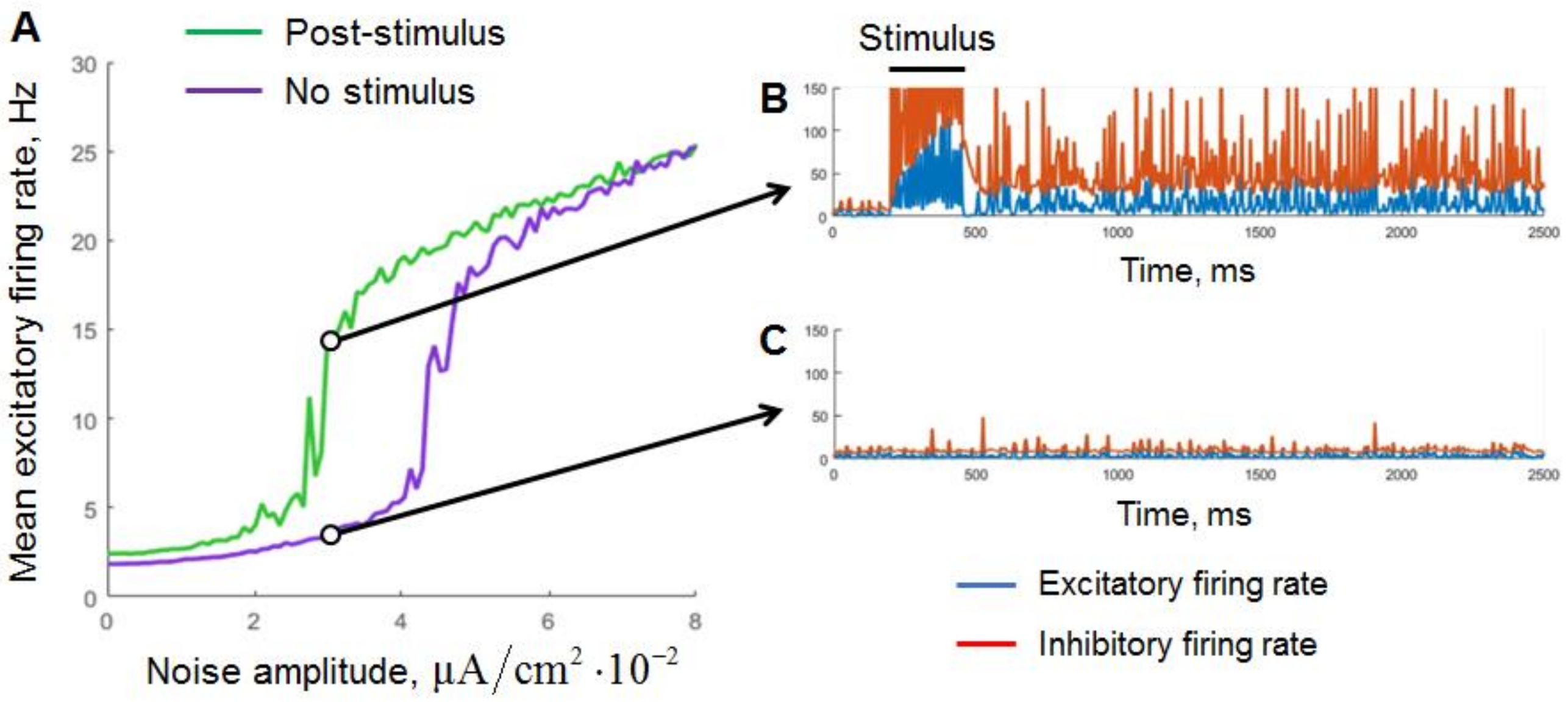

Figure 4. Stabilization of the single-circuit metastable activity by a noisy input. (A) Mean excitatory firing rate, averaged over the interval 1 – 10 seconds. Green – the result for post-stimulus activity; purple – for background activity (without stimulus presentation). The stimulus presentation interval was 200 – 450 ms. The noise was delivered during the whole simulation time. Examples of the post-stimulus and the background

activity traces ( $A_{ex,NOISE} = 3 \cdot 10^{-2}\ \mu\text{A}/\text{cm}^2$ ) are presented in (B) and (C), respectively. Blue curves – excitatory firing rate, red curves – inhibitory firing rate.

### 3.2 Activity of two-circuit model

In this section, we explore joint stabilizing effect of input noise and oscillations on the system of two circuits with mutual excitatory connections (Fig. 1B). We investigate different cases when circuits receive in-phase or anti-phase oscillations, as well as common or independent noise. The inter-circuit connections contain both fast AMPA and slow NMDA components. We varied the proportion of the inter-circuit current that flows via NMDA receptors (which we denoted as $k_{NMDA}^{cross}$), while keeping the total inter-circuit connection strength at the constant level.

Dependence of the post-stimulus activity duration (which we use as the measure of the active regime stability) on $k_{NMDA}^{cross}$ value and the amplitude of the input oscillations is presented in Fig. 5. In the case of fast (AMPA) inter-circuit connections ($k_{NMDA}^{cross} = 0$), the in-phase oscillatory input to both circuits leads to stabilization of the active regime with high efficiency (Fig. 5, lower part of the left panel). Anti-phase input oscillations in this case need to have a very high amplitude to stabilize the active regime (Fig. 5, lower part of the right panel). With increasing $k_{NMDA}^{cross}$ (i.e. the portion of the slow NMDA component of the inter-circuit current), effectiveness of the in-phase input deteriorates, while effectiveness of the anti-phase input increases (see the opposite trends in the left and right panels of Fig. 5). For $k_{NMDA}^{cross} \cong 1$, both types of the input have approximately the same effect (Fig 5, upper parts of the panels).

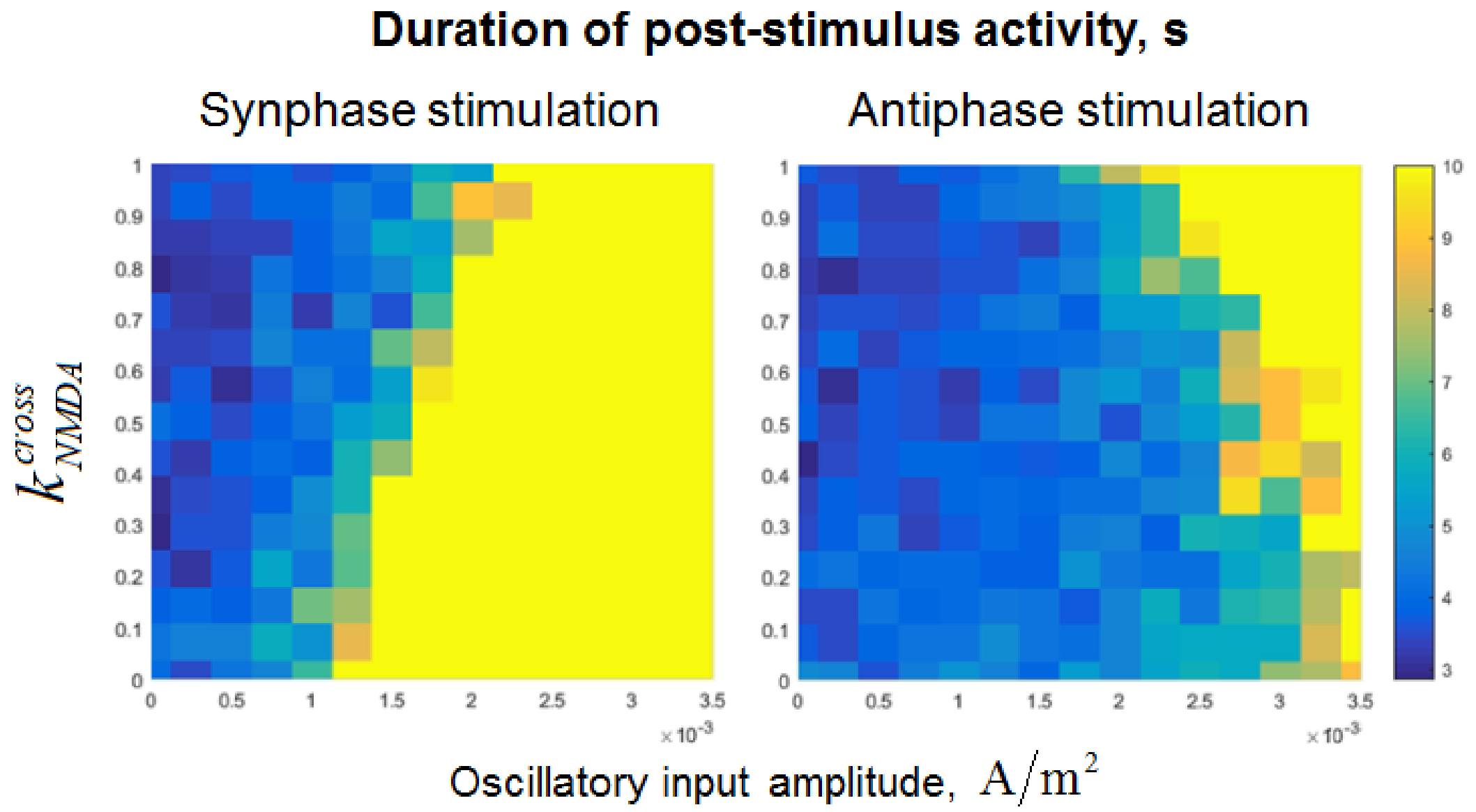

Figure 5. Dependence of post-stimulus activity duration of the two-circuit system on the amplitude of input sinusoidal oscillations and the NMDA-to-total inter-circuit current ratio ( $k_{NMDA}^{cross}$ ). Left panel: the oscillations are delivered to the circuits in the same phase, right panel: oscillations are delivered in the antiphase. Oscillation frequency: 40 Hz, noise amplitude: 0.014 $\mu A/cm^2$. Note that increasing of $k_{NMDA}^{cross}$ reduces the stabilizing effect of the in-phase input, but increases the effect of the antiphase input.

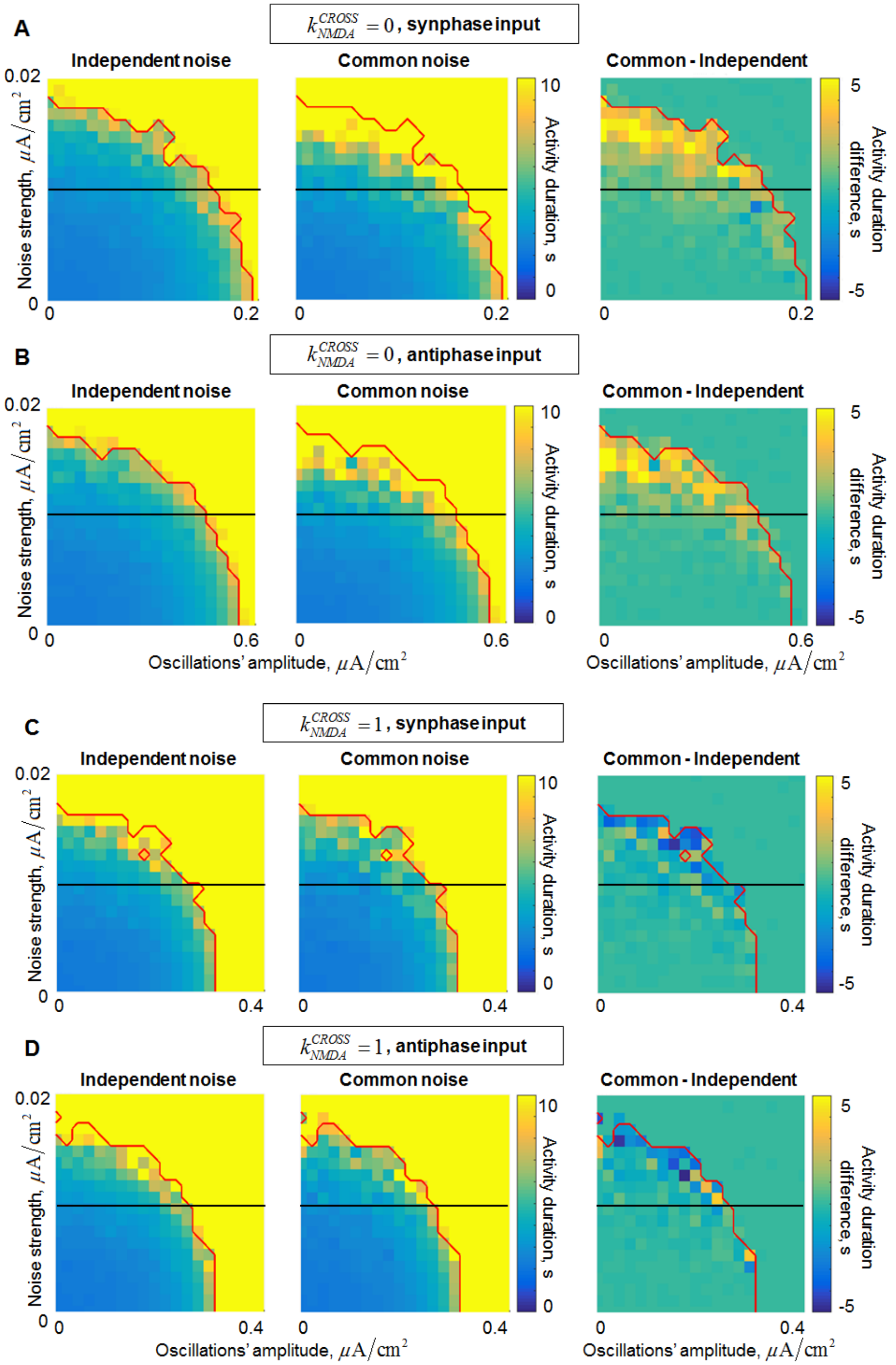

Figure 6. Dependence of the post-stimulus activity duration in the two-circuit system on the amplitude of the input oscillations and the strength of the input noise for the in/anti-phase oscillations and two values of the NMDA connections strength. (A,B) Fast inter-circuit connection, (C,D) slow inter-circuit connections. (A,C) The circuits receive oscillatory signals in the same phase, (B,D) the circuits receive oscillatory signals in the opposite phases. Left column – each circuit receives an independent noise input; middle column – the circuits

receive a common noise input; right column – the difference between these two cases. Red curve denotes the border of the saturation region: above this curve, the activity duration either for the common or for the independent noise case equals to the simulation time (i.e., above this curve comparison between the noise types does not make sense). Horizontal black lines denote the noise level of 0.01 (used further in the text). When the inter-circuit connections are fast, the activity is more robust (its duration is longer) in the case of common noise (yellow regions in the right panels of (A,B)). When the connections are slow, the activity is slightly more robust in the case of independent noise (blue regions in the right panels of (C,D)).

In the presence of noise, our system generates irregular gamma-band quasi-oscillations. They act together with the external gamma-band input in stabilizing the active regime. The joint dependence of the post-stimulus activity duration on the input oscillations' amplitude and on the noise standard deviation is presented in Fig. 6. In general, the activity duration increases when either the oscillations or the noise become stronger.

In the case of fast inter-circuit connections ($k_{NMDA}^{cross} = 0$, Figs. 6A,B), common noise has stronger stabilizing effect than independent noise (compare the left and middle panels of Figs. 6A,B). This is true in the presence of either in-phase or anti-phase oscillations, but in the case of anti-phase oscillations, the stabilizing effect occurs at much higher amplitudes compared to in-phase oscillations (see different horizontal scales in Figs. 6A and 6B). The yellow region in the right panels of Figs. 6A,B contains the combinations of oscillations' amplitude and noise standard deviation, for which there is an evident difference in the post-stimulus activity duration between the common-noise and the independent-noise cases (within our simulation time). Importantly, for intermediate noise intensities, an evident difference is observed only if the oscillations are sufficiently strong (see the horizontal black lines in the right panels of Figs. 6A,B cross the yellow region near the right parts of the diagrams). In other words, an external gamma-band input with an appropriate amplitude stabilizes the active regime (i.e. WM retention) only when the two circuits receive a common noise input (e.g. when they belong to the same distributed representation), but not when they receive independent noise inputs of the same strength (e.g. when they are parts of different representations).

In the case of slow inter-circuit connections ($k_{NMDA}^{cross} = 1$, Figs. 6C,D), common noise has weaker stabilizing effect than independent noise (compare the left and middle panels of Figs. 6A,B). This difference, however, is less pronounced than in the case of fast inter-circuit connections (compare the right panels of Figs. 6A,B and 6C,D). Furthermore, in-phase and antiphase gamma-band inputs have almost the same effect (compare Figs. 6C and 6D, note that the horizontal scale is the same). This is in contrast with the $k_{NMDA}^{cross} = 0$ case, in which in-phase input had much stronger effect than the antiphase input. Such difference between $k_{NMDA}^{cross} = 0$ and $k_{NMDA}^{cross} = 1$ is in agreement with the result presented in Fig. 5 for a constant noise intensity. Let us, again, select a certain intermediate noise intensity (horizontal black lines in Figs. 6C,D), and start to increase the gamma-band input amplitude. When the amplitude becomes high enough, the gamma-band input begins to stabilize the activity regime. This happens at slightly smaller amplitudes if the noise is independent, but irrespectively of whether the oscillatory input is in-phase or antiphase.

### 3.3 Oscillatory control of multi-circuit system

In this section, we show how the behaviour of a multi-circuit system can be controlled by various types of oscillatory inputs. The results presented here are, in a large part, based on the effects of oscillations on two-circuit models described in the previous section. The multi-circuit system under consideration contains two clusters of circuits. Each cluster contains a group of circuits receiving a common noise input (the same signal for both clusters) and a group of circuits receiving independent noise inputs (we denote the "common-noise" groups of the first and second cluster as C1 and C2, respectively, and the “independent-noise” groups as I1 and I2) . We consider two models – with fast and slow inter-cluster connections, respectively. We explore the behavior of the models in four conditions: (1) no oscillatory input (“NONE” condition), (2) only one cluster receives oscillatory input (“1CLUST” condition), (3) the clusters receive in-phase oscillations (“SYNC” condition), (4) the clusters receive anti-phase oscillations (“ANTI” condition). Each simulation was performed 25 times, and the statistics of group-averaged post-stimulus activity duration were collected for each of the four groups (C1, C2, I1, I2) separately.

Fig. 7 summarizes the results on the multi-circuit system behavior under the various oscillatory input conditions. The range of post-stimulus activity durations of the “common-noise” groups (C1, C2) is represented by vertical red bars; the range of activity durations of the “independent-noise” groups (I1, I2) – by vertical black bars. The horizontal dash in the middle of a bar marks the corresponding median value. Medians of the black and red bars obtained in the same condition are connected by black lines; a slope of such line demonstrates the difference in activity durations between the “common-noise” and “independent-noise” groups.

Without an oscillatory input (“NONE” condition), the groups receiving common noise input (C1, C2) stay active for slightly longer time than the groups receiving independent noise inputs (I1, I2). When the oscillatory input is delivered to the first cluster (“1CLUST” condition), it increases post-stimulus activity duration of both groups that belong to this cluster (C1, I1). Importantly, the activity duration difference between the “common-noise” group (C1) and the “independent- noise” group (I1) also increases. These effects are almost absent for the second cluster (groups C2, I2) due to relatively weak inter-cluster interaction.

When the oscillations are delivered to both clusters, either in the same phase or in the antiphase (“SYNC” and “ANTI” conditions, respectively), they produce similar effects to one-cluster oscillatory input, but these effects are stronger and involve both clusters. Thus, post-stimulus activity duration for all the groups (C1, I1, C2, I2), as well as activity duration difference between the “common-noise” and “independent-noise” groups (C1-I1, C2-I2) are increased. In Fig. 7, we can see that the activity duration is higher in “SYNC” and “ANTI” conditions, compared to “NONE” and “1CLUST” conditions for all noise input types (red and black), clusters (1 and 2) and model types (slow and fast inter-cluster connections). It is also seen that the links between the red and black boxes in both clusters (C1-I1, C2-I2) are steeper in “SYNC” / “ANTI” conditions than in “NONE” / “1CLUST” conditions (for both model types), which reflects increased duration difference between the “common-noise” and “independent-noise” groups.

The models with fast and slow inter-cluster connections differ in their response to in-phase and anti-phase oscillatory inputs. In the model with fast connections, the in-phase input provides strong increase both in the activity duration and in the duration difference between the “common-noise” and “independent-noise” groups, while both these effects are considerably weaker in the case of the anti-phase input. On the contrary, in the model with slow inter-cluster connections, the in-phase and anti-phase oscillatory inputs lead to the effects of roughly the same strength. This difference between the two models is best visible in the results averaged over the clusters (right panels in Figs. 7A,B). It is seen that the median levels and the slopes of the “red-black” links are higher for “SYNC” condition than for “ANTI” condition in the right panel of Fig. 7A. At the same time, both the median levels and the slopes for “SYNC” and “ANTI” conditions are close to each other Fig. 7B.

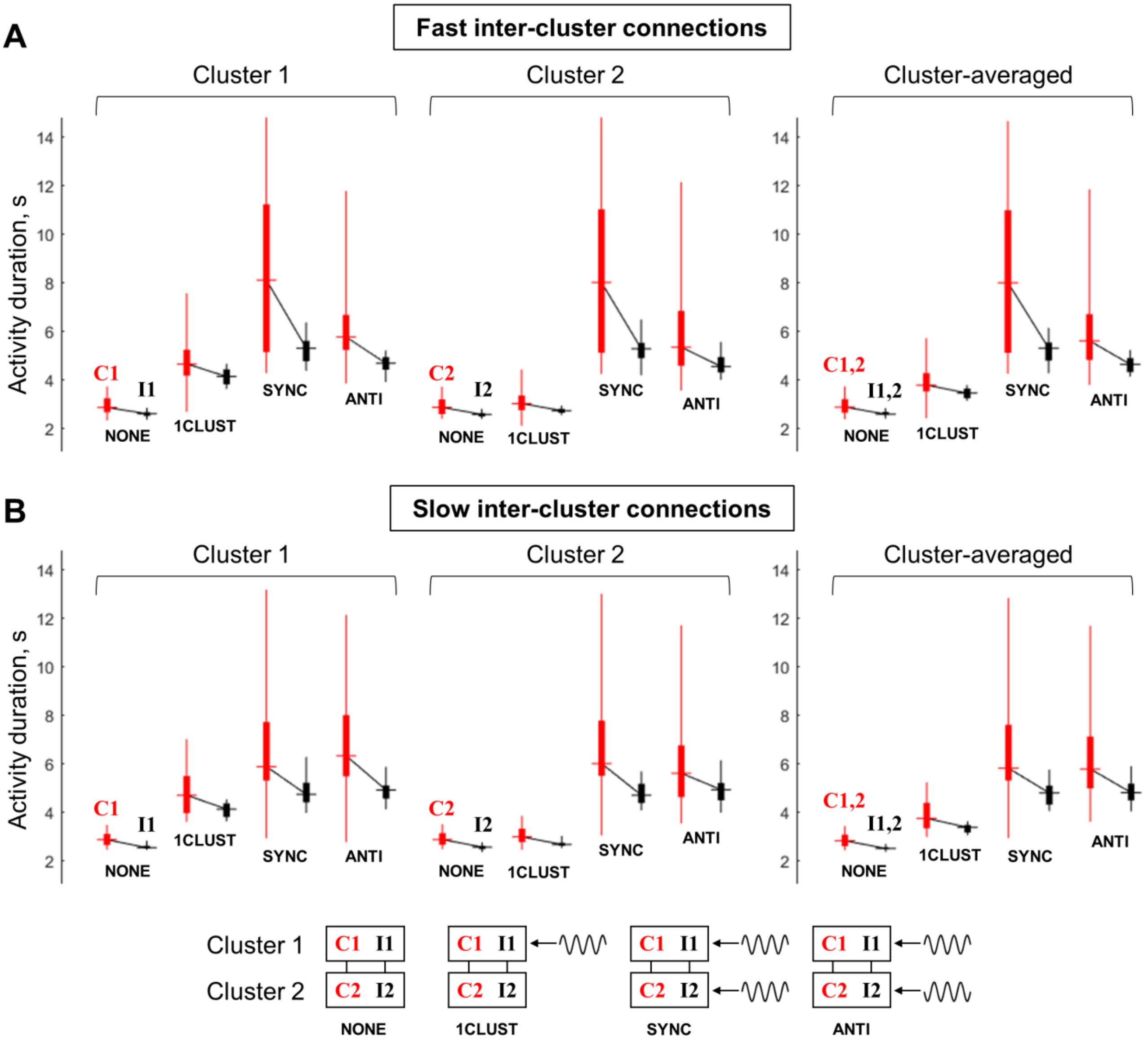

Figure 7. Duration of post-stimulus activity for various circuit groups in the multi-circuit model, depending on the type of oscillatory input. A – the model with fast inter-cluster connections, B – the model with slow inter-cluster connections. The connected pairs of red and black vertical bars correspond to pairs of circuit groups within the same cluster. Red bars correspond to the groups that receive a common noise input (C1, C2), black bars – to the groups that receive independent noise inputs (I1, I2). Each bar represents statistics of group-averaged post-stimulus activity durations (obtained in 25 simulation runs): horizontal dash – the median, thick vertical line – two middle quartiles, thin vertical line – range between the minimal and maximal values. Left panel: cluster 1 (groups C1, I1), middle panel: cluster 2 (groups C2, I2), right panel – average over the two clusters (C1+C2, I1+I2). Four pairs of boxes in each panel correspond to four types of the oscillatory input: NONE – no oscillations, 1CLUST – oscillations are delivered to one cluster only, SYNC - oscillations are delivered to both clusters in the same phase, ANTI - oscillations are delivered to both clusters in the opposite phases (schematic representations on the input types are presented in the bottom part of the figure). It is seen that oscillatory input increases the activity duration, as well as the difference in activity duration between the "common-noise" and "independent-noise" groups; these effects are stronger when oscillations are delivered to both clusters. The effects of the in-phase input are more prominent than of the anti-phase input for the model with fast inter-cluster connections. For the model with slow inter-cluster connections, the in-phase and anti-phase inputs produce the effects of the same strength.

## 4 Discussion

In this study, we considered several WM model variants, comprising of one, two, or many excitatory-inhibitory metastable circuits having mutual excitatory connections and receiving the same stimulus-related signal. First, we explored a single-circuit system and demonstrated that input gamma oscillations or white noise could stabilize (i.e. prolong) post-stimulus metastable activity. Next, we considered a two-circuit model and found that: (1) fast (AMPA) inter-circuit connections provide better stabilization by common-noise input compared to independent-noise inputs, (2) this difference could be further amplified by oscillatory inputs, (3) in-phase oscillatory inputs to the circuits produce better stabilization than anti-phase inputs in the case of fast (AMPA) connections, (4) in-phase and anti-phase inputs produce similar effects in the case of slow (NMDA) connections.

Finally, we developed a more realistic, multi-circuit system able to align with the effects observed in the two-circuit model. The system comprised of two clusters receiving in-phase or anti-phase oscillations and linked by fast (AMPA) or slow (NMDA) connections. Each cluster contained a group of circuits that receive a common noise input (representing activity of a distributed WM representation this group is embedded into) and a group of circuits that receive independent inputs (and, thus, do not participate in the distributed WM retention). We found that: (1) oscillatory inputs stabilize activity of all circuits, (2) this stabilization is more pronounced for the "common-noise" groups compared to “independent-noise” groups, (3) in the case of fast inter-cluster connections, both the stabilization and separation between the "common-noise" and “independent-noise” groups are more pronounced for in-phase input compared to anti-phase input, (4) in the case of slow connections, in-phase and anti-phase inputs produce comparable effects.

In summary, we demonstrated that gamma oscillations are able to selectively stabilize activity of the circuits that receive common noise input, thus supporting coherent activity of a distributed WM representation. If long-range connections are fast, a part of this representation could be further highlighted by synchronizing gamma activity within this part. If these connections are slow, the distributed representation is uniformly stabilized, irrespective of phase differences between gamma oscillations in its local parts.

### Single-circuit system

We started from a single-circuit system and demonstrated that, with an appropriate parameter selection, both the noise and the gamma-band input alone could stabilize WM retention, without affecting the background regime. To explain these effects, we note that our model was tuned to operate in a subthreshold regime, in which the gain functions of the populations (excitatory and inhibitory) were concave. Consequently, even a zero-mean input (noisy or oscillatory) produced additional mean excitation to the populations. In our system, the mean effect of oscillations on the excitatory population outweighed their effect on the inhibitory population, which resulted in oscillation-induced / noise-induced mean firing rate increase and, consequently, to stabilization of the active regime.

A similar effect – stabilization of initially metastable active regime in a WM model by gamma-band input – was previously demonstrated by Schmidt et al. (2018). In this study, the authors considered a purely excitatory system, but took into account the effect of spike-to-spike synchronization, which allowed to achieve resonant behavior without inhibitory population. The system had a resonance in the beta band, but the stabilization occurred under high-gamma input, while beta-band input, on the contrary, switched the system to the background state. A similar switching to the background state by a resonant input was also reported by Dipoppa and Gutkin (2013). In our model, we do not observe this effect, presumably due to the presence of slow NMDA currents ant short-term plasticity, which make the activity more robust to transient episodes of synchronization. As a result, the effect of input oscillations is always excitatory in our model.

We should note that our model is more biologically realistic, compared to Schmidt et al. (2018) and Dipoppa, Gutkin (2013). First, it contains both an excitatory and an inhibitory population. Second, it operates in a subthreshold regime (which is typical for cortical networks during WM retention, (Compte et al., 2003)), while in the aforementioned models, the neurons operated mostly in a suprathreshold regime with regular spiking activity. Third, both these models used highly non-linear pulse-like periodic inputs, while in our case, the oscillatory input was sinusoidal (which is, again, closer to the situation in the neocortex; see (Wang, 2010)).

**Two-circuit system**

In the two-circuit system, under a constant noise intensity, we demonstrated that in-phase gamma-band input stabilizes WM retention much more effectively than antiphase input when inter-circuit interaction is fully mediated by fast AMPA receptors ($k_{NMDA}^{cross} = 0$). With increasing NMDA:AMPA ratio of the inter-circuit connections, effectiveness of the in-phase gamma-band input decreased, while effectiveness of the antiphase input increased. When the interaction was fully mediated by slow NMDA receptors ($k_{NMDA}^{cross} = 1$), effectiveness of in-phase and antiphase input was about the same.

The observed effects could be explained by the fact that the eigenmode of the system is in-phase if $k_{NMDA}^{cross} = 0$ and antiphase if $k_{NMDA}^{cross} = 1$. This fact is illustrated in Figs. S2E and S3E. Without oscillatory input, independent noise inputs lead to in-phase quasi-oscillations if $k_{NMDA}^{cross} = 0$ (left part of Fig. S2E, black curve); however, if $k_{NMDA}^{cross} = 1$, then the noise-induced quasi-oscillations have the average phase difference between the circuits closer to π (left part of Fig. S3E, black curve). The most effective entrainment (and, thus, the most effective stabilization) occurs when the phase difference between the inputs to the circuits matches the eigenmode of the system. In the case of $k_{NMDA}^{cross} = 0$, antiphase input should change activity of the system from the in-phase to the antiphase mode to fully entrain it (Fig. S2F), which is accompanied by oscillation-induced desynchronization at certain input amplitudes (Fig. S2D); in the case of $k_{NMDA}^{cross} = 1$, the antiphase entrainment is much easier (Fig. S3D,F). On the contrary, in-phase oscillations more easily entrain the system if $k_{NMDA}^{cross} = 0$ (Fig. S2C,E), rather than if $k_{NMDA}^{cross} = 1$ (Fig. S3C,E).

We next considered how the duration of post-stimulus activity (i.e. effectiveness of the active regime stabilization) jointly depends on noise intensity and gamma-band input amplitude. For the fast inter-circuit connections ($k_{NMDA}^{cross} = 0$), common noise stabilized the active regime more effectively than independent noise. On the contrary, for the slow inter-circuit connections ($k_{NMDA}^{cross} = 1$), independent noise was slightly more effective. Importantly, for intermediate noise intensities, there was a range of gamma amplitudes, in which the gamma-band input stabilized the active regime with considerably different effectiveness depending on the noise type (common / independent). We used this effect as the basis for selective oscillatory control of activity robustness in the multi-circuit system.

The higher effectiveness of the common noise in the case of $k_{NMDA}^{cross} = 0$ could be, again, explained by the fact that the system's eigenmode in this case is in-phase. Common noise fully projects to this mode, while independent noise projects to it only partially. Thus, entrainment of the in-phase mode by independent noise is weaker, which leads to smaller stabilizing effect. On the contrary, independent noise partially projects to the anti-phase mode (which is the eigenmode of the system with $k_{NMDA}^{cross} = 1$), while common noise is orthogonal to this mode, so independent noise is more effective when $k_{NMDA}^{cross} = 1$.

There is another effect that possibly participates in the link between $k_{NMDA}^{cross}$ and activity robustness. Since slow NMDA receptors provide low-pass filtering of activity, they make the system less resonant. Consequently, increase of $k_{NMDA}^{cross}$ decreases the amplitude of entrained oscillations or noise-induced quasi-oscillations (compare left parts of Figs. S2G, S3G). This effect presumably sums up with the aforementioned effects of eigenmode matching by external inputs.

We note that worsening of the stabilization by in-phase oscillations for higher NMDA:AMPA ratio seemingly contradicts earlier results suggesting the stabilizing role of NMDA receptors in WM retention in the presence of oscillatory activity (Tegner, et al., 2002). This discrepancy could be explained by the fact that the model used in (Tegner, et al., 2002) is bistable, but does not contain any slow variables except of the NMDA current. Consequently, the system in the active state is stable in the absence of fluctuations, but fast fluctuations (such as gamma oscillations) make it fall down to the background state if the NMDA:AMPA ratio is too small. On the contrary, our model is metastable, so it spontaneously returns from the active regime to the background state in the absence of fluctuations. At the same time, our system contains slow variables besides the inter-circuit NMDA current – namely, within-circuit NMDA currents and coefficients of within-circuit short-term plasticity. As the result, our system is resistant to fast fluctuations (i.e. it survives fluctuation-induced "gaps" in the activity), even if the within-circuit NMDA:AMPA ratio is zero. Fast fluctuations, instead, increase the level of activity (thus stabilizing it) due to concavity of the gain functions. Notably, the ability of periodic input to stabilize metastable regime even in a model without slow variables was previously demonstrated by Schmidt et al. (2018). We suggest that the stabilizing effect of NMDA current described in (Tegner, et al., 2002) could be also present in our model, but it is weak compared to the other effects we described.

**Multi-circuit system**

In the analysis of the two-circuit system, we observed two important effects. First, in the case of fast connections and intermediate noise level, input oscillations could strongly stabilize WM retention in a pair of circuits receiving a common noise input, while producing much weaker effect on circuits receiving independent noise inputs. Second, if the connections between two circuits are fast, then the stabilizing effect of oscillations is considerably stronger when oscillations are delivered to the circuits in the same phase rather than in the opposite phases; however, this difference is mitigated if the inter-circuit connections are slow.

In order to demonstrate how these effects could be scaled up and utilized for oscillatory control of WM retention, we developed a multi-circuit system. We assumed that WM retention is based on distributed neural activity, and that local cortical patches could contain circuits involved in this activity, as well as circuits not involved in it (a similar scheme was used in (Lundqvist et al., 2011), in which local modules contained parts of various representations, and corresponding parts in different modules were connected by long-range projections). We explicitly included two local patches (referred to as clusters) into our WM system and modelled the inputs from other parts of the cortex as white-noise signals. The circuits in each cluster formed two groups, named C1, I1 in the first cluster, and C2, I2 – in the second one. The groups C1, C2 were considered to participate in the coherent distributed activity related to WM retention, so the circuits from these groups received common noise input. The groups I1, I2 did not participate in this activity, so their circuits received independent noise inputs. We note that, unlike most WM models (e.g. Brunel, Wang, 2001; Lundqvist et al., 2011), participation of a group in the WM representation is not determined at the stimulus presentation stage, since all the circuits in our model receive the same stimulus signal. Instead, a group is considered to be a part of the WM representation if its circuits receive common noise input.

We demonstrated that, as in the two-circuit case, gamma-band input can stabilize WM retention, i.e. increase post-stimulus activity duration. Importantly, this stabilization was more prominent for the “common-noise” groups (C1, C2), compared to the “independent-noise” groups (I1, I2). Thus, the gamma-band input also increased “selectivity” of WM retention, predominantly stabilizing those circuits that participate in WM-related distributed coherent activity. This behavior is in agreement with the reported increase in both the firing rates and stimulus selectivity that occurs on top of elevated gamma activity episodes during WM retention (Lundqvist et al., 2016, 2018; Bastos et al., 2018). Such functionality of gamma-input in our model is achieved due to high AMPA-based within-group connectivity and low inter-group connectivity. As a consequence of this, circuit pairs with common noise input and pairs with independent inputs are mostly separated from each other, so the results obtained for the two-circuit system are still applicable. If the inter-group connections are too dense, correlations would spread across the whole system, and the difference between the “common-noise” and “independent-noise” circuits would be diminished.

We suggest that the dense within-group and sparse inter-group connectivity follows naturally from the Hebbian plasticity principles: the circuits that receive the same input (members of C1, C2) would instantiate links between each other. We assume that the “independent-input” circuits (members of I1, I2) could also participate in the dominant active representation (and thus receive a common input) in certain cases that we do not model explicitly here: e.g. for different WM content or different task rules. On the contrary, since the groups C1 and I1 (as well as C2 and I2) participate in different representations, their circuits do not usually receive common inputs, which justifies sparse inter-group (C1 – I1, C2 – I2) connections.

We also confirmed in the multi-circuit module the influence of NMDA:AMPA ratio on the system’s tendency to respond differently depending on the phase between gamma-band oscillatory inputs delivered to its parts. We demonstrated that the stabilizing effect of the input oscillations (i.e. oscillation-induced prolongation of post-stimulus activity) in the system with fast (AMPA-based) inter-cluster connections is stronger when the oscillations are delivered to the clusters in the same phase rather than in the opposite phases. On the contrary, the stabilizing effect of the in-phase and anti-phase inputs to the clusters was the same in the case of slow (NMDA-based) inter-cluster connections. This difference between the slow ant fast inter-cluster connections was observed only when the connectivity between the corresponding groups of different clusters (C1 – C2, I1 – I2) was weaker than the within-group connectivity. We suggest that such layout is biologically plausible, due to presumed small-world character of the cortical topology (Bullmore, Sporns, 2009; Bassett, Bullmore, 2017), with long-range connections being, in general, sparser than short-range connections (Ercsey-Ravasz et al., 2013).

Thus, fast and slow inter-cluster connections provide different functionality. Fast connections allow to selectively increase retention robustness in a part of an active distributed representation by synchronizing it in the gamma-band (e.g., by providing common gamma-band input from a controller cortical or subcortical network). In turn, slow inter-cluster connections provide a basis for robust WM retention by a distributed network with local gamma generators, which do not need to be synchronized. We note that the local character of gamma oscillations is experimentally supported (Donner, Siegel, 2011), and it was also assumed in previous multi-modular models of WM (Lundqvist et al., 2011).

**Funding**

This article is an output of a research project implemented as part of the Basic Research Program at the National Research University Higher School of Economics (HSE University). BG acknowledges support from CNRS, INSERM, ANR-17-EURE-0017, and ANR-10-IDEX-0001-02.

**Conflict of Interest**

The authors declare that the research was conducted in the absence of any commercial or financial relationships that could be construed as a potential conflict of interest.

**References**

Amit, D. J., and Brunel, N. (1997). Model of global spontaneous activity and local structured activity during delay periods in the cerebral cortex. *Cereb. Cortex* 7, 237–252. doi:10.1093/cercor/7.3.237.

Ardid, S., Wang, X. J., Gomez-Cabrero, D., and Compte, A. (2010). Reconciling coherent oscillation with modulation of irregular spiking activity in selective attention: Gamma-range synchronization between sensory and executive cortical areas. *J. Neurosci.* 30, 2856–2870. doi:10.1523/JNEUROSCI.4222-09.2010.

Augustin, M., Ladenbauer, J., Baumann, F., and Obermayer, K. (2017). Low-dimensional spike rate models derived from networks of adaptive integrate-and-fire neurons: Comparison and implementation. *PLoS Comput Biol.*, 13(6), e1005545. doi:10.1371/journal.pcbi.1005545

Baddeley, A. (2003). Working memory: Looking back and looking forward. *Nat. Rev. Neurosci.* 4, 829–839. doi:10.1038/nrn1201.

Bassett, D. S., and Bullmore, E. T. (2017). Small-World Brain Networks Revisited. *Neuroscientist*. *23*(5), 499–516. doi:10.1177/1073858416667720Bastos, A. M., Loonis, R., Kornblith, S., Lundqvist, M., and Miller, E. K. (2018). Laminar recordings in frontal cortex suggest distinct layers for maintenance and control of working memory. *Proc. Natl. Acad. Sci. U. S. A.* 115, 1117–1122. doi:10.1073/pnas.1710323115.

Brunel, N., and Wang, X. J. (2001). Effects of neuromodulation in a cortical network model of object working memory dominated by recurrent inhibition. *J. Comput. Neurosci.* 11, 63–85. doi:10.1023/a:1011204814320.

Bullmore, E., and Sporns, O. (2009). Complex brain networks: graph theoretical analysis of structural and functional systems. *Nat Rev Neurosci*. 10(3), 186–198. doi:10.1038/nrn2575

Chafee, M. V., and Goldman-Rakic, P. S. (1998). Matching patterns of activity in primate prefrontal area 8a and parietal area 7ip neurons during a spatial working memory task. *J. Neurophysiol.* 79, 2919–2940. doi:10.1152/jn.1998.79.6.2919.

Chik, D. (2013). Theta-alpha cross-frequency synchronization facilitates working memory control - A modeling study. *Springerplus* 2, 1–10. doi:10.1186/2193-1801-2-14.

Compte, A., Constantinidis, C., Tegnér, J., Raghavachari, S., Chafee, M. V., Goldman-Rakic, P. S., et al. (2003). Temporally Irregular Mnemonic Persistent Activity in Prefrontal Neurons of Monkeys during a Delayed Response Task. *J. Neurophysiol.* 90, 3441–3454. doi:10.1152/jn.00949.2002.

Compte A. (2006). Computational and in vitro studies of persistent activity: edging towards cellular and synaptic mechanisms of working memory. Neuroscience. 139(1), 135–151. doi: 10.1016/j.neuroscience.2005.06.011

Constantinidis, C., and Goldman-Rakic, P. S. (2002). Correlated discharges among putative pyramidal neurons and interneurons in the primate prefrontal cortex. *J. Neurophysiol.* 88, 3487–3497. doi:10.1152/jn.00188.2002.

Dipoppa, M., and Gutkin, B. S. (2013). Flexible frequency control of cortical oscillations enables computations required for working memory. *Proc. Natl. Acad. Sci. U. S. A.* 110, 12828–12833. doi:10.1073/pnas.1303270110.

Donner, T. H., and Siegel, M. (2011). A framework for local cortical oscillation patterns. *Trends Cogn Sci*. 15(5), 191–199. doi:10.1016/j.tics.2011.03.007

Ercsey-Ravasz, M., Markov, N. T., Lamy, C., Van Essen, D. C., Knoblauch, K., Toroczkai, Z., and Kennedy, H. (2013). A predictive network model of cerebral cortical connectivity based on a distance rule. *Neuron. 80*(1), 184–197. doi:10.1016/j.neuron.2013.07.036

Funahashi, S., Bruce, C. J., and Goldman-Rakic, P. S. (1989). Mnemonic coding of visual space in the monkey’s dorsolateral prefrontal cortex. *J. Neurophysiol.* 61, 331–349. doi:10.1152/jn.1989.61.2.331.

Fuster, J. M., and Alexander, G. E. (1971). Neuron activity related to short-term memory. *Science (80-. ).* 173, 652–654. Available at: http://www.ncbi.nlm.nih.gov/pubmed/4998337.

Goldman-Rakic, P. S. (1995). Cellular basis of working memory. *Neuron* 14, 477–485. Available at: http://www.ncbi.nlm.nih.gov/pubmed/7695894.

Gutkin, B. S., Laing, C. R., Colby, C. L., Chow, C. C., and Ermentrout, G. B. (2001). Turning on and off with excitation: The role of spike-timing asynchrony and synchrony in sustained neural activity. *J. Comput. Neurosci.* 11, 121–134. doi:10.1023/A:1012837415096.

Haegens, S., Osipova, D., Oostenveld, R., and Jensen, O. (2010). Somatosensory working memory performance in humans depends on both engagement and disengagement of regions in a distributed network. *Hum. Brain Mapp.* 31, 26–35. doi:10.1002/hbm.20842.

Hansel, D., and Mato, G. (2013). Short-term plasticity explains irregular persistent activity in working memory tasks. *J. Neurosci.* 33, 133–149. doi:10.1523/JNEUROSCI.3455-12.2013.

Haegens, S., Osipova, D., Oostenveld, R., and Jensen, O. (2010). Somatosensory working memory performance in humans depends on both engagement and disengagement of regions in a distributed network. *Hum. Brain Mapp.* 31, 26–35. doi:10.1002/hbm.20842.

Howard, M. W., Rizzuto, D. S., Caplan, J. B., Madsen, J. R., Lisman, J., Aschenbrenner-Scheibe, R., Schulze-Bonhage, A., & Kahana, M. J. (2003). Gamma oscillations correlate with working memory load in humans. Cereb Cortex. 13(12), 1369–1374. doi: 10.1093/cercor/bhg084

Jokisch, D., and Jensen, O. (2007). Modulation of gamma and alpha activity during a working memory task engaging the dorsal or ventral stream. *J. Neurosci.* 27, 3244–3251. doi:10.1523/JNEUROSCI.5399-06.2007.

Kaiser, J., Ripper, B., Birbaumer, N., and Lutzenberger, W. (2003). Dynamics of gamma-band activity in human magnetoencephalogram during auditory pattern working memory. *Neuroimage* 20, 816–827. doi:10.1016/S1053-8119(03)00350-1.

Kopell, N., Whittington, M. A., and Kramer, M. A. (2011). Neuronal assembly dynamics in the beta1 frequency range permits short-term memory. *Proc. Natl. Acad. Sci. U. S. A.* 108, 3779–3784. doi:10.1073/pnas.1019676108.

Kornblith, S., Buschman, T. J., and Miller, E. K. (2016). Stimulus load and oscillatory activity in higher cortex. *Cereb. Cortex* 26, 3772–3784. doi:10.1093/cercor/bhv182.

Laing, C. R., and Chow, C. C. (2001). Stationary bumps in networks of spiking neurons. *Neural Comput.* 13, 1473–1494. doi:10.1162/089976601750264974.

Liebe, S., Hoerzer, G. M., Logothetis, N. K., and Rainer, G. (2012). Theta coupling between V4 and prefrontal cortex predicts visual short-term memory performance. *Nat. Neurosci.* 15, 456–462. doi:10.1038/nn.3038.

Lim, S., and Goldman, M. S. (2013). Balanced cortical microcircuitry for maintaining information in working memory. *Nat. Neurosci.* 16, 1306–1314. doi:10.1038/nn.3492.

Lisman, J. E., and Idiart, M. A. P. (1995). Storage of 7 ± 2 short-term memories in oscillatory subcycles. *Science (80-. ).* 267, 1512–1515. doi:10.1126/science.7878473.

Lundqvist, M., Compte, A., and Lansner, A. (2010). Bistable, irregular firing and population oscillations in a modular attractor memory network. *PLoS Comput. Biol.* 6, 1–12. doi:10.1371/journal.pcbi.1000803.

Lundqvist, M., Herman, P., and Lansner, A. (2011). Theta and gamma power increases and alpha/beta power decreases with memory load in an attractor network model. *J. Cogn. Neurosci.* 23, 3008–3020. doi:10.1162/jocn_a_00029.

Lundqvist, M., Herman, P., Warden, M. R., Brincat, S. L., and Miller, E. K. (2018). Gamma and beta bursts during working memory readout suggest roles in its volitional control. *Nat. Commun.* 9. doi:10.1038/s41467-017-02791-8.

Lundqvist, M., Rose, J., Herman, P., Brincat, S. L. L., Buschman, T. J. J., and Miller, E. K. K. (2016). Gamma and Beta Bursts Underlie Working Memory. *Neuron* 90, 152–164. doi:10.1016/j.neuron.2016.02.028.

Lutzenberger, W., Ripper, B., Busse, L., Birbaumer, N., and Kaiser, J. (2002). Dynamics of gamma-band activity during an audiospatial working memory task in humans. *J. Neurosci.* 22, 5630–5638. doi:10.1523/JNEUROSCI.22-13-05630.2002.

Miller, E. K., Erickson, C. A., and Desimone, R. (1996). Neural mechanisms of visual working memory in prefrontal cortex of the macaque. *J. Neurosci.* 16, 5154–5167. doi:10.1523/JNEUROSCI.16-16-05154.1996.

Mongillo, G., Barak, O., and Tsodyks, M. (2008). Synaptic theory of working memory. *Science* 319, 1543–1546. doi:10.1126/science.1150769.

Mongillo, G., Hansel, D., and van Vreeswijk, C. (2012). Bistability and spatiotemporal irregularity in neuronal networks with nonlinear synaptic transmission. *Phys. Rev. Lett.* 108, 158101. doi:10.1103/PhysRevLett.108.158101.

Palva, J. M., Monto, S., Kulashekhar, S., and Palva, S. (2010). Neuronal synchrony reveals working memory networks and predicts individual memory capacity. *Proc. Natl. Acad. Sci. U. S. A.* 107, 7580–7585. doi:10.1073/pnas.0913113107.

Palva, S., Kulashekhar, S., Hämäläinen, M., and Palva, J. M. (2011). Localization of cortical phase and amplitude dynamics during visual working memory encoding and retention. *J. Neurosci.* 31, 5013–5025. doi:10.1523/JNEUROSCI.5592-10.2011.

Pina, J. E., Bodner, M., and Ermentrout, B. (2018). Oscillations in working memory and neural binding: A mechanism for multiple memories and their interactions. *PLoS Comput. Biol.* 14. doi:10.1371/journal.pcbi.1006517.

Renart, A., Moreno-Bote, R., Wang, X. J., and Parga, N. (2007). Mean-driven and fluctuation-driven persistent activity in recurrent networks. *Neural Comput.*, 19(1), 1–46. doi:10.1162/neco.2007.19.1.1

Roux, F., and Uhlhaas, P. J. (2014). Working memory and neural oscillations: α-γ versus θ-γ codes for distinct WM information? *Trends Cogn. Sci.* 18, 16–25. doi:10.1016/j.tics.2013.10.010.

Roxin, A., and Compte, A. (2016). Oscillations in the bistable regime of neuronal networks. *Phys. Rev. E* 94. doi:10.1103/PhysRevE.94.012410.

Sauseng, P., Klimesch, W., Heise, K. F., Gruber, W. R., Holz, E., Karim, A. A., et al. (2009). Brain Oscillatory Substrates of Visual Short-Term Memory Capacity. *Curr. Biol.* 19, 1846–1852. doi:10.1016/j.cub.2009.08.062.

Schaffer, E. S., Ostojic, S., and Abbott, L. F. (2013). A complex-valued firing-rate model that approximates the dynamics of spiking networks. *PLoS Comput Biol.*, *9*(10), e1003301. doi:10.1371/journal.pcbi.1003301

Schmidt, H., Avitabile, D., Montbrió, E., and Roxin, A. (2018). Network mechanisms underlying the role of oscillations in cognitive tasks. *PLoS Comput. Biol.* 14. doi:10.1371/journal.pcbi.1006430.

Sherfey, J., Ardid, S., Miller, E. K., Hasselmo, M. E., and Kopell, N. J. (2020). Prefrontal oscillations modulate the propagation of neuronal activity required for working memory. *Neurobiol. Learn. Mem.* 173. doi:10.1016/j.nlm.2020.107228.

Siegel, M., Warden, M. R., and Miller, E. K. (2009). Phase-dependent neuronal coding of objects in short-term memory. *Proc. Natl. Acad. Sci. U. S. A.* 106, 21341–21346. doi:10.1073/pnas.0908193106.

Tegnér, J., Compte, A., and Wang, X. J. (2002). The dynamical stability of reverberatory neural circuits. *Biol. Cybern.* 87, 471–481. doi:10.1007/s00422-002-0363-9.

Tseng, P., Chang, Y.-T., Chang, C.-F., Liang, W.-K., and Juan, C.-H. (2016). The critical role of phase difference in gamma oscillation within the temporoparietal network for binding visual working memory. *Sci. Rep.* 6, 32138. doi:10.1038/srep32138.

Tsodyks, M. V, and Markram, H. (1997). The neural code between neocortical pyramidal neurons depends on neurotransmitter release probability. *Proc. Natl. Acad. Sci. U. S. A.* 94, 719–723. doi:10.1073/pnas.94.2.719.

van Vugt, M. K., Schulze-Bonhage, A., Litt, B., Brandt, A., and Kahana, M. J. (2010). Hippocampal gamma oscillations increase with memory load. *J. Neurosci.* 30, 2694–2699. doi:10.1523/JNEUROSCI.0567-09.2010.

Wang, X. J. (2001). Synaptic reverberation underlying mnemonic persistent activity. *Trends Neurosci.* 24, 455–463. doi:10.1016/s0166-2236(00)01868-3.

Wang, X. J. (2010). Neurophysiological and computational principles of cortical rhythms in cognition. *Physiol. Rev.* 90, 1195–1268. doi:10.1152/physrev.00035.2008.

Wimmer, K., Ramon, M., Pasternak, T., and Compte, A. (2016). Transitions between multiband oscillatory patterns characterize memory-guided perceptual decisions in prefrontal circuits. *J. Neurosci.* 36, 489–505. doi:10.1523/JNEUROSCI.3678-15.2016.

# Supplementary Material

## 5 Model

A generic multi-circuit model is described by the following equations:

$$
\begin{cases}
\tau_{ra}\dfrac{dr_a^p}{dt} = -r_a^p + F_{ra}\left(\mu_a^p, \sigma_{a,AMPA}^p, \sigma_{a,GABAA}^p\right) \\
\mu_a^p = \mu_{a,AMPA}^p + \mu_{a,NMDA}^p + \mu_{a,GABAA}^p \\
\tau_{AMPA}\dfrac{d\mu_{a,AMPA}^p}{dt} = -\mu_{a,AMPA}^p + \sum_q \tilde{J}_{ae,AMPA}^{pq} K_{ae}\tau_{AMPA} r_e^q + \mu_{ax,BG}^p + \mu_{ax,NOISE}^p + \mu_{ax,STIM}^p + \mu_{ax,OSC}^p \\
\tau_{NMDA}\dfrac{d\mu_{a,NMDA}^p}{dt} = -\mu_{a,NMDA}^p + \sum_q \tilde{J}_{ae,NMDA}^{pq} K_{ae}\tau_{NMDA} r_e^q \\
\tau_{GABAA}\dfrac{d\mu_{a,GABAA}^p}{dt} = -\mu_{a,GABAA}^p + \sum_q \tilde{J}_{ai,GABAA}^{pq} K_{ai}\tau_{GABAA} r_i^q \\
\dfrac{\tau_{AMPA}}{2}\dfrac{d\left(\sigma_{a,AMPA}^p\right)^2}{dt} = -\left(\sigma_{a,AMPA}^p\right)^2 + \dfrac{1}{2}\sum_q \left(\tilde{J}_{ae,AMPA}^{pq}\right)^2 K_{ae}\tau_{AMPA} r_e^q + \left(\sigma_{ax,BG}^p\right)^2 \\
\sigma_{a,NMDA}^p = 0 \\
\dfrac{\tau_{GABAA}}{2}\dfrac{d\left(\sigma_{a,GABAA}^p\right)^2}{dt} = -\left(\sigma_{a,GABAA}^p\right)^2 + \dfrac{1}{2}\sum_q \left(\tilde{J}_{ai,GABAA}^{pq}\right)^2 K_{ai}\tau_{GABAA} r_i^q \\
\mu_{ax,NOISE}^p(t) = A_{ax,NOISE}^p \cdot \xi_{ax}^p(t) \\
\mu_{ax,STIM}^p(t) = A_{ax,STIM}^p \cdot \left[t_{STIM,1} \le t \le t_{STIM,2}\right] \\
\mu_{ax,OSC}^p(t) = A_{ax,OSC}^p \cdot \sin\left(2\pi f_{OSC}\cdot\left(t - t_{OSC,1}\right) + \varphi_{ax,OSC}^p\right)\cdot\left[t_{OSC,1} \le t \le t_{OSC,2}\right] \\
\dfrac{du^p}{dt} = -\dfrac{u^p}{\tau_F} + U(1-u^p) r_e^p \\
\dfrac{dx^p}{dt} = -\dfrac{(1-x^p)}{\tau_D} - u^p x^p r_e^p \\
\tilde{J}_{ab,S}^{pq} = \begin{cases} J_{ab,S}^{pq} x^p u^p, & \text{if } p = q,\ ab = ee,\ S \in \{AMPA, NMDA\} \\ J_{ab,S}^{pq}, & \text{otherwise} \end{cases}
\end{cases}
\tag{1}
$$

where the lower indexes *a,b* correspond to the population types (*e* – excitatory, *i* – inhibitory), whereas the upper indexes *p,q* are the number of a circuit. The variable $r_a^p$ is the firing rate of a population. The variable $\mu_a^p$ is the average input current, and $\mu_{a,AMPA}^p, \mu_{a,NMDA}^p, \mu_{a,GABAA}^p$ are the components of this current corresponding to the AMPA, NMDA ,and GABA synapses respectively; $\sigma_{a,AMPA}^p, \sigma_{a,NMDA}^p, \sigma_{a,GABAA}^p$ are the standard deviations of these current components (taken over the neurons of a population). The variables $x^p, u^p$ describe the synaptic short-term plasticity. $F_{ra}$ is the neuronal transfer function, $\tau_{ra}$ is the time constant of the population firing rate dynamics, and $\tau_{AMPA}, \tau_{NMDA}, \tau_{GABAA}$ are the synaptic time constants. $K_{ab}$ is the number of synaptic projections from the population *b* to the population *a*. $J_{ae,AMPA}^{pq}, J_{ae,NMDA}^{pq}$ are the weights of the AMPA and NMDA connections from the excitatory population of the circuit $q$ to the population $a$ of the circuit $p$. $J_{ai,GABAA}^{pq}$ are the weights of the connections from the inhibitory population of the circuit $q$ to the population $a$ of the circuit $p$; $\tilde{J}_{ab,S}^{pq}$ $\left(S \in \{AMPA, NMDA, GABAA\}\right)$ are the weights of the connections with the short term plasticity taken into account. The variables $\mu_{ax,BG}^p$, $\sigma_{ax,BG}^p$ are the population average and standard deviation of the external background input from other brain areas that were not explicitly included into the model; $\mu_{ax,STIM}^p(t)$, $\mu_{ax,NOISE}^p(t)$, $\mu_{ax,OSC}^p(t)$ are the external input signals corresponding to the stimulus, the noise and the oscillations respectively. $A_{ax,STIM}^p$ is the stimulus amplitude, $t_{STIM,1}$ and $t_{STIM,2}$ are the start and end time of the stimulus; $A_{ax,NOISE}^p$ is the noise standard deviation, $\xi_{ax}^p(t)$ is the Gaussian white noise with the zero average and unit variance, $A_{ax,OSC}^p, f_{OSC}$ are the amplitude and frequency of the input oscillations, $t_{OSC,1}, t_{OSC,2}$ are the start and end time of the oscillatory input; $\tau_F, \tau_D$ are the time constant of synaptic potentiation and depression respectively; $U$ is the coefficient affecting the ratio between the potentiation and the depression.

For parameterization of the system, it is convenient to set the total level of excitation transmitted through the AMPA and NMDA receptors, as well as the ratio between the AMPA and NMDA components. Parameterization is also simplified by identity of the circuits. Thus, within-circuit connections could be described as follows:

$$\begin{cases} J_{ae,AMPA}^{pp} = J_{ae}(1-k_{NMDA}) \\ J_{ae,NMDA}^{pp} = J_{ae}k_{NMDA}\,\tau_{AMPA}/\tau_{NMDA}\,, \\ J_{ai,GABAA}^{pp} = J_{ai} \end{cases} \qquad (2)$$

where $J_{ae}, J_{ai}$ are the variables that characterize excitation and inhibition within the circuits; $k_{NMDA}$ is a coefficient showing the contribution of the NMDA synapses in within-circuit excitation.

For the two-circuit model, we describe the inter-circuit weights as follows:

$$\begin{cases} J_{ee,AMPA}^{12} = J_{ee,AMPA}^{21} = J_{ee}^{cross}(1-k_{NMDA}^{cross}) \\ J_{ee,NMDA}^{12} = J_{ee,NMDA}^{21} = J_{ee}^{cross}(1-k_{NMDA}^{cross})\,\tau_{AMPA}/\tau_{NMDA} \qquad , \\ J_{ab,S}^{12} = J_{ab,S}^{21} = 0, \quad ab \neq ee,\ S \in \{AMPA, NMDA, GABAA\} \end{cases} \qquad (3)$$

where $J_{ee}^{cross}$ is the strength of the circuit interaction; $k_{NMDA}^{cross}$ is a coefficient showing the contribution of NMDA current in circuit interaction.

Our multi-circuit model contains of two clusters. Each cluster contains two groups of 8 circuits each (which we denote as C1, I1 for the cluster 1 and C2, I2 for the cluster 2). The circuits from the first groups of the clusters (C1, C2) receive a common noise input, and the circuits from the second groups (I1, I2) – independent noise inputs.

For the multi-circuit model, we separately define the weights and NMDA coefficients for within-cluster and inter-cluster connections:

$$\begin{cases} \text{For } p,q \text{ from the same cluster:} \\ J_{ee,AMPA}^{pq} = c_{pq} J_{ee}^{clust} (1 - k_{NMDA}^{clust}) \\ J_{ee,NMDA}^{pq} = c_{pq} J_{ee}^{clust} (1 - k_{NMDA}^{clust}) \tau_{AMPA} / \tau_{NMDA} \\ \text{For } p,q \text{ from different clusters:} \\ J_{ee,AMPA}^{pq} = c_{pq} J_{ee}^{far} (1 - k_{NMDA}^{far}) \\ J_{ee,NMDA}^{pq} = c_{pq} J_{ee}^{far} (1 - k_{NMDA}^{far}) \tau_{AMPA} / \tau_{NMDA} \\ \text{For } all \;\; p \neq q: \\ J_{ab,S}^{pq} = 0, \;\; ab \neq ee, \; S \in \{AMPA, NMDA, GABAA\} \end{cases}, \qquad (4)$$

where $J_{ee}^{clust}, J_{ee}^{far}$ are the strengths of within-cluster and inter-cluster connections, respectively; $k_{NMDA}^{clust}, k_{NMDA}^{far}$ are coefficients showing the contributions of the NMDA current in within-cluster and inter-cluster connections, respectively; $c_{pq}$ is a binary coefficient showing presence or absence of a connection between circuits $p$ and $q$. All the connections are symmetric: $c_{pq} = c_{qp}$.

The connectivity pattern $\{c_{pq}\}$ is chosen randomly according to the following rules:

- All the connections are symmetric: $c_{pq} = c_{qp}$;
- Each circuit within a group (C1, C2, I1, I2) receives 3 inputs from circuits of the same group;
- Different groups of the same cluster (C1-I1 and C2-I2) have 3 connections between each other, each circuit has either zero or one inter-group connection;
- Same groups of the different clusters (C1-C2, I1-I2) have 8 connections with each other, each circuit of a group has exactly one inter-cluster connection;
- Different groups of different clusters (C1-I2, C2-I1) are not connected.

Parameters of the tonic input, noise and external stimulus are the same for all circuits:

$$\begin{cases} \mu_{ax,BG}^{p} \equiv \mu_{ax,BG} \\ \sigma_{ax,BG}^{p} \equiv \sigma_{ax,BG} \\ A_{ax,NOISE}^{p} \equiv A_{ax,NOISE} \\ A_{ax,STIM}^{p} \equiv A_{ax,STIM} \end{cases}. \qquad (5)$$

External oscillations are delivered to the excitatory populations of the circuits only. Amplitude of oscillations is the same for all circuits.

$$\begin{cases} A^{p}_{ex,OSC} \equiv A_{ex,OSC} \\ A^{p}_{ix,OSC} = 0 \end{cases}. \tag{6}$$

We consider either in-phase or anti-phase oscillatory inputs to circuits. In the two-circuit model, the circuits could receive oscillations in the same phase or in the opposite phases:

$$\begin{cases} \varphi^{1}_{ex,OSC} = \varphi^{1}_{ix,OSC} = \varphi^{2}_{ix,OSC} = 0 \\ \varphi^{2}_{ex,OSC} = \begin{bmatrix} 0, \text{ synphase} \\ \pi, \text{ antiphase} \end{bmatrix} \end{cases} \tag{7}$$

In the multi-circuit model, all circuits from a cluster receive an input of the same phase, while the phases could be the same or the opposite for the two clusters. In our notations:

$$\begin{cases} \varphi^{p}_{ix,OSC} = 0, \text{ for any } p \\ \varphi^{p}_{ex,OSC} = 0, \text{ for } p \text{ from the cluster 1} \\ \varphi^{p}_{ex,OSC} = \begin{bmatrix} 0, \text{ synphase} \\ \pi, \text{ antiphase} \end{bmatrix}, \text{ for } p \text{ from the cluster 2} \end{cases} \tag{8}$$

The noise input is delivered to the excitatory populations of the circuits only, with the same amplitude for all circuits:

$$\begin{cases} A^{p}_{ex,NOISE} \equiv A_{ex,NOISE} \\ A^{p}_{ix,NOISE} = 0 \end{cases}. \tag{9}$$

In the two-circuit model, the noise inputs to the circuits could be common ($\xi^{1}_{ex}(t) = \xi^{2}_{ex}(t)$) or independently generated. In the multi-circuit model, the noise input is identical for all the circuits that belong to the groups C1, C2 ($\xi^{p}_{ex}(t) = \xi^{COMMON}_{ex}(t)$), while each circuit from the groups I1, I2 receives independently generated noise.

The model parameters were set in such a way as to ensure metastability and resonant properties of the active regime. First, we selected parameters that ensure bistability in a single circuit (Fig. 2B). In this case, a pair of complex-conjugate eigenvalues for the active state should correspond to slowly decaying oscillations in the gamma band (see Fig. S1B,D for example orbits). After that, the excitatory-to excitatory connection weight was decreased until the active state lost stability (Fig. 2A). Next, we considered two circuits and connected them with an excitatory connection, simultaneously decreasing the tonic inputs to the excitatory populations, in order to compensate for the inter-circuit excitation.

In the multi-circuit models, the inter-circuit weights were further decreased (compared to the two-circuit model), given the increased number of connections per circuit. The inter-cluster weights were made smaller that the within-cluster weights to keep the clusters relatively independent from each other. The proportion between the number of inter-group and within-group connections was made small enough, so the circuits receiving common noise could profit from it; otherwise, input correlations would spread across the whole system due to inter-circuit connectivity, compromising the difference between the “common-noise” and “independent-noise” groups. The noise intensity was set just enough for the activity duration difference between the “common-noise” and “independent-noise” groups to become visible. Finally, the oscillatory input amplitude was selected such that the oscillations considerably increase the activity duration and the difference between the “common-noise” and “independent-noise” groups both in the case of fast and slow inter-cluster connections.

The values of the model parameters are provided in Table 1.

Numerical simulations of the models were carried out using a self-developed software package in the Matlab environment. Numerical integration was performed using the Euler method with 0.2 ms time steps for the single-circuit and two-circuit models, and with 0.5 ms steps for the multi-circuit models.

Table 1. Model parameters

| **Common parameters** | | | |
|---|---|---|---|
| **Parameter** | **Value** | **Parameter** | **Value** |
| $\tau_{re}$ | 3 ms | $K_e$ | 400 |
| $\tau_{ri}$ | 1.5 ms | $K_i$ | 100 |
| $\tau_e^{AMPA}$ | 2 ms | $J_{ee}$ | 2.8 μA/cm$^2$ |
| $\tau_e^{NMDA}$ | 50 ms | $J_{ie}$ | 0.29 μA/cm$^2$ |
| $\tau_i^{GABAA}$ | 5 ms | $k_{NMDA}$ | 0.7 |
| $g_m$ | 100 $\mu S/cm^2$ | $J_{ei}$ | -0.15 μA/cm$^2$ |
| $C_{me}$ | 2 $\mu F/cm^2$ | $J_{ii}$ | -0.09 μA/cm$^2$ |
| $C_{mi}$ | 1 $\mu F/cm^2$ | $\mu_{ix,BG}$ | 0.25 μA/cm$^2$ |
| $E_L$ | -70 mV | $\sigma_{ex,BG}$ | 0.02 μA/cm$^2$ |
| $V_t$ | -50 mV | $\sigma_{ix,BG}$ | 0.02 μA/cm$^2$ |
| $V_r$ | -60 mV | $\tau_F$ | 450 ms |
| $U$ | 0.03 | $\tau_D$ | 200 ms |
| **Single-circuit model** | | | |

| Parameter | Value | Parameter | Value |
|---|---|---|---|
| $\mu_{ex,BG}$ | 1 μA/cm$^2$ | $A_{ex,NOISE}$ | Control parameter |
| $A_{ex,STIM}$ | 5 μA/cm$^2$ | $f_{OSC}$ | Control parameter |
| $A_{ix,STIM}$ | 1 μA/cm$^2$ | $A_{ex,OSC}$ | Control parameter |
| $t_{STIM,1}$ | 200 ms | $\Delta\varphi_{OSC}$ | 0 or π |
| $t_{STIM,2}$ | 450 ms | $t_{OSC,1}$ | 800 ms |
| | | $t_{OSC,2}$ | Inf |
| **Two-circuit model** | | | |
| **Parameter** | **Value** | **Parameter** | **Value** |
| $\mu_{ex,BG}$ | 0.8 μA/cm$^2$ | $A_{ex,NOISE}$ | 0.014 μA/cm$^2$ |
| $A_{ex,STIM}$ | 5 μA/cm$^2$ | $f_{OSC}$ | 40 Hz |
| $A_{ix,STIM}$ | 1 μA/cm$^2$ | $A_{ex,OSC}$ | Control parameter |
| $t_{STIM,1}$ | 200 ms | $\Delta\varphi_{OSC}$ | 0 or π |
| $t_{STIM,2}$ | 450 ms | $t_{OSC,1}$ | 800 ms |
| $J_{ee}^{cross}$ | 0.026 μA/cm$^2$ | $t_{OSC,2}$ | Inf |
| $k_{NMDA}^{cross}$ | Control parameter | | |
| **Multi-circuit model** | | | |
| **Parameter** | **Value** | **Parameter** | **Value** |
| $\mu_{ex,BG}$ | 0.8 μA/cm$^2$ | $A_{ex,NOISE}$ | 0.02 μA/cm$^2$ |
| $A_{ex,STIM}$ | 3.5 μA/cm$^2$ | $f_{OSC}$ | 40 Hz |
| $A_{ix,STIM}$ | 0.7 μA/cm$^2$ | $A_{ex,OSC}$ | 0.18 μA/cm$^2$ |
| $t_{STIM,1}$ | 100 ms | $\Delta\varphi_{OSC}$ (within-cluster) | 0 |
| $t_{STIM,2}$ | 450 ms | $\Delta\varphi_{OSC}$ (inter-cluster) | 0 or π |
| $J_{ee}^{clust}$ | 0.015 μA/cm$^2$ | $t_{OSC,1}$ | 550 ms |
| $k_{NMDA}^{clust}$ | 0 | $t_{OSC,2}$ | Inf |
| $J_{ee}^{far}$ | 0.003 μA/cm$^2$ | | |
| $k_{NMDA}^{far}$ | 0.1 or 1 | | |

## 6 Calculation of the transfer functions

The neuronal transfer functions relate the synaptic input to the value of the output firing rate. In our case, the arguments of the transfer function are the average synaptic current $\mu$ and the standard deviations of the excitatory AMPA-current $\sigma_{AMPA}$ and the inhibitory GABAA-current $\sigma_{GABAA}$. To determine the value of the transfer function for a certain combination $(\mu, \sigma_{AMPA}, \sigma_{GABAA})$ we performed a simulation of a single leaky integrate-and-fire (LIF) neuron that receives a tonic current input $\mu$ and two noisy inputs with the zero average, standard deviations $\sigma_{AMPA}, \sigma_{GABAA}$, and time constants $\tau_{AMPA}, \tau_{GABAA}$:

$$\begin{cases} C_{ma}\dfrac{dV}{dt} = g_{La}(\mu - V) + I_{AMPA} + I_{GABAA} \\ if\ V = V_a^{TH} : V \leftarrow V_a^R \\ \tau_{AMPA}\dfrac{dI_{AMPA}}{dt} = -I_{AMPA} + \sigma_{AMPA}\sqrt{2\tau_{AMPA}}\eta_{AMPA}(t) \\ \tau_{GABAA}\dfrac{dI_{GABAA}}{dt} = -I_{GABAA} + \sigma_{GABAA}\sqrt{2\tau_{GABAA}}\eta_{GABAA}(t) \end{cases}, \qquad (10)$$

where $C_{ma}$ is the membrane capacity, $g_{La}$ is its conductivity, $V_a^{TH}$ is a spike generation threshold, $V_a^R$ is a reset voltage, $I_{AMPA}, I_{GABAA}$ are the membrane currents through the AMPA- and GABAA-receptors respectively, and $\eta_{AMPA}(t)$, $\eta_{GABAA}(t)$ are independent implementations of Gaussian white noise with zero mean and unit variance. The parameter set is presented in the Table 2.

We carry out a simulation of the system (10) for different combinations $(\mu, \sigma_{AMPA}, \sigma_{GABAA})$, located in the nodes of a rectangular grid, and for each combination we stored the resulting firing rate. This procedure was performed twice: for an excitatory and for an inhibitory neuron. Thus, two three-dimensional arrays $\hat{F}_{re}, \hat{F}_{ri}$ were obtained. In the process of simulating the population model (1) and analyzing the phase plane, we determined the values of the functions $F_{re}, F_{ri}$ at the required points using cubic interpolation between the precomputed values $\hat{F}_{re}, \hat{F}_{ri}$ at the nearest grid nodes.

The shapes of the resulting transfer functions $F_{re}(\mu, \sigma_{AMPA}, \sigma_{GABAA})$ and $F_{ri}(\mu, \sigma_{AMPA}, \sigma_{GABAA})$ as functions of $\mu$, given constant values of $\sigma_{AMPA}, \sigma_{GABAA}$, are presented in Figs. 2C,D, respectively. For this demonstration, we considered again a bistable single-circuit model (obtained by increasing $J_{ee}$). We plotted the shapes of $F_{re}$ and $F_{ri}$ as functions of $\mu$ for two different combinations of $\sigma_{AMPA}, \sigma_{GABAA}$ – one corresponding to the background state (solid curves in Figs. 2C,D) and the other – to the active state (dashed curves in Figs. 2C,D).

Table 2. The parameter set for computing of the transfer functions

| Parameter | Value | Parameter | Value |
|---|---|---|---|
| $C_{me}$ | 2 μF/cm[2] | $E_{Le} = E_{Li}$ | -70 mV |
| $C_{mi}$ | 1 μF/cm[2] | $\tau_{AMPA}$ | 2 ms |
| $g_{Le} = g_{Li}$ | 100 μS/cm[2] | $\tau_{NMDA}$ | 50 ms |
| $V_e^{TH} = V_i^{TH}$ | -50 mV | $\tau_{GABAA}$ | 5 ms |
| $V_e^R = V_i^R$ | -60 mV | | |

In our model, we assume that NMDA current variance is negligible and do not use it as an additional argument of the transfer functions. To justify it, we consider the following ratio:

$$\frac{\sigma_{NMDA}}{\sigma_{AMPA}} = \frac{J_{NMDA}\sqrt{K_e r_e \tau_{NMDA}}}{J_{AMPA}\sqrt{K_e r_e \tau_{AMPA}}} = \frac{J_{NMDA} K_e r_e \tau_{NMDA}}{J_{AMPA} K_e r_e \tau_{AMPA}} \sqrt{\frac{\tau_{AMPA}}{\tau_{NMDA}}} = \frac{\mu_{AMPA}}{\mu_{NMDA}} \sqrt{\frac{\tau_{AMPA}}{\tau_{NMDA}}}. \qquad (11)$$

If we assume that the mean currents are of the same order of magnitude ( $\mu_{AMPA}/\mu_{NMDA} \sim 1$) and the NMDA time constant is much larger than the AMPA time constant ( $\tau_{AMPA}/\tau_{NMDA} \approx 0$ ), then we conclude from (11) that $\sigma_{AMPA}/\sigma_{NMDA} \approx 0$ .

## 7 Phase portrait analysis

For a deeper understanding of the system's behavior, we visualized a phase plane $(r_e, r_i)$ with the characteristic curves (Figs. 2A,B, S1). For each combination $(r_e, r_i)$, we determined the values of all other variables, at which their time derivatives turn to zero:

$$\begin{cases} u^*(r_e) = U r_e \tau_F / (1 + U r_e \tau_F) \\ x^*(r_e) = 1 / (1 + u^*(r_e) r_e \tau_D) \\ G_a^{syn}(r_e) = \begin{bmatrix} x^*(r_e) u^*(r_e), & a = e \\ 1, & a = i \end{bmatrix} \\ \mu^*_{a,AMPA}(r_e, r_i) = J_{ae,AMPA} G_a^{syn}(r_e) K_{ae} \tau_{AMPA} r_e + \mu_{ax,BG} \\ \mu^*_{a,NMDA}(r_e, r_i) = J_{ae,NMDA} G_a^{syn}(r_e) K_{ae} \tau_{NMDA} r_e \\ \mu^*_{a,GABAA}(r_e, r_i) = J_{ai,GABAA} K_{ai} \tau_{GABAA} r_i \\ \left(\sigma^*_{a,AMPA}\right)^2 (r_e, r_i) = \frac{1}{2} J^2_{ae,AMPA} \left(G_a^{syn}(r_e)\right)^2 K_{ae} \tau_{AMPA} r_e + \sigma^2_{ax,BG} \\ \left(\sigma^*_{a,GABAA}\right)^2 (r_e, r_i) = \frac{1}{2} J^2_{ai,GABAA} K_{ai} \tau_{GABAA} r_i \end{cases} . \qquad (12)$$

We defined $r_e$ -curve as the set of points on the plane $(r_e, r_i)$ for which the time derivative of $r_e$ is zero, taking into account the conditions (12). Similarly, we defined $r_i$- curve as the set of points for which the time derivative of $r_i$ is zero, and the conditions (12) are satisfied. Thus, we obtained the following expressions for both curves:

$$\begin{cases} r_e = F_{re}\left(\mu_e^*(r_e,r_i),\sigma_{e,AMPA}^*(r_e,r_i),\sigma_{e,GABAA}^*(r_e,r_i)\right) \\ r_i = F_{ri}\left(\mu_i^*(r_e,r_i),\sigma_{i,AMPA}^*(r_e,r_i),\sigma_{i,GABAA}^*(r_e,r_i)\right) \end{cases}. \quad (13)$$

The intersections of the curves (two for a bistable system, one for metastable systems) correspond to the steady-states.

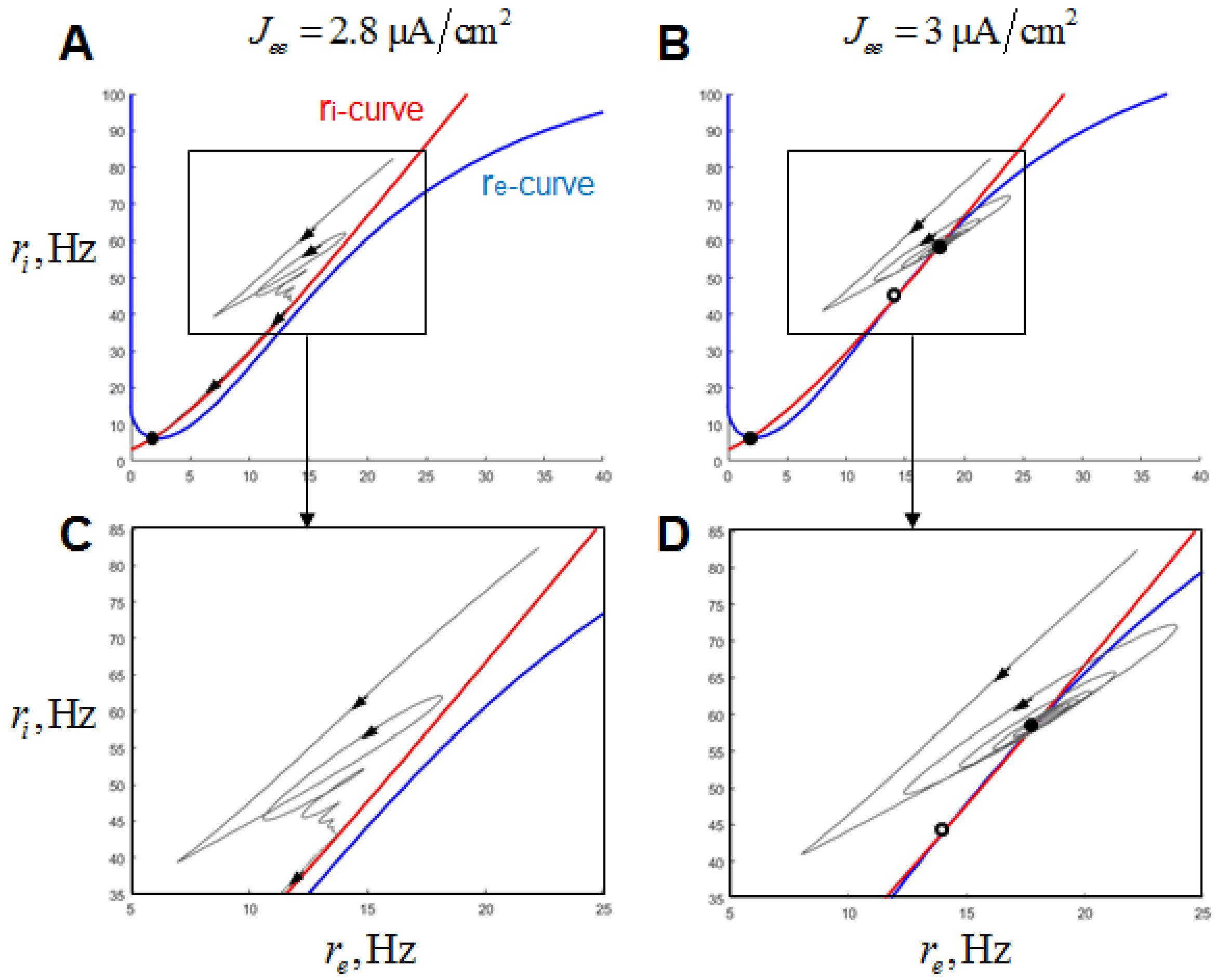

Figure S1. Damped oscillations on a phase plane. The legend is the same as in Fig. 2. (A, C) The system we analyzed in the paper, with the single steady-state (background) and the region of slowly decaying activity (where the re- and ri-curves are close to each other). The presented orbit approaches ri-curve in the region of high firing rates, demonstrating damped oscillations, and then goes towards the background steady state, staying close to ri-curve. (B, D) The system with increased recurrent excitation, with two steady-states (background and active). The presented orbit demonstrates damped oscillations around a point close to the active state and the converges to the active state itself. Since the presented phase plane is a projection of multi-dimensional state space, the orbits have self-intersections. For both systems, kNMDA was increased from the original value 0.7 to 0.73, for illustrative purposes (to increase the damping ratio).

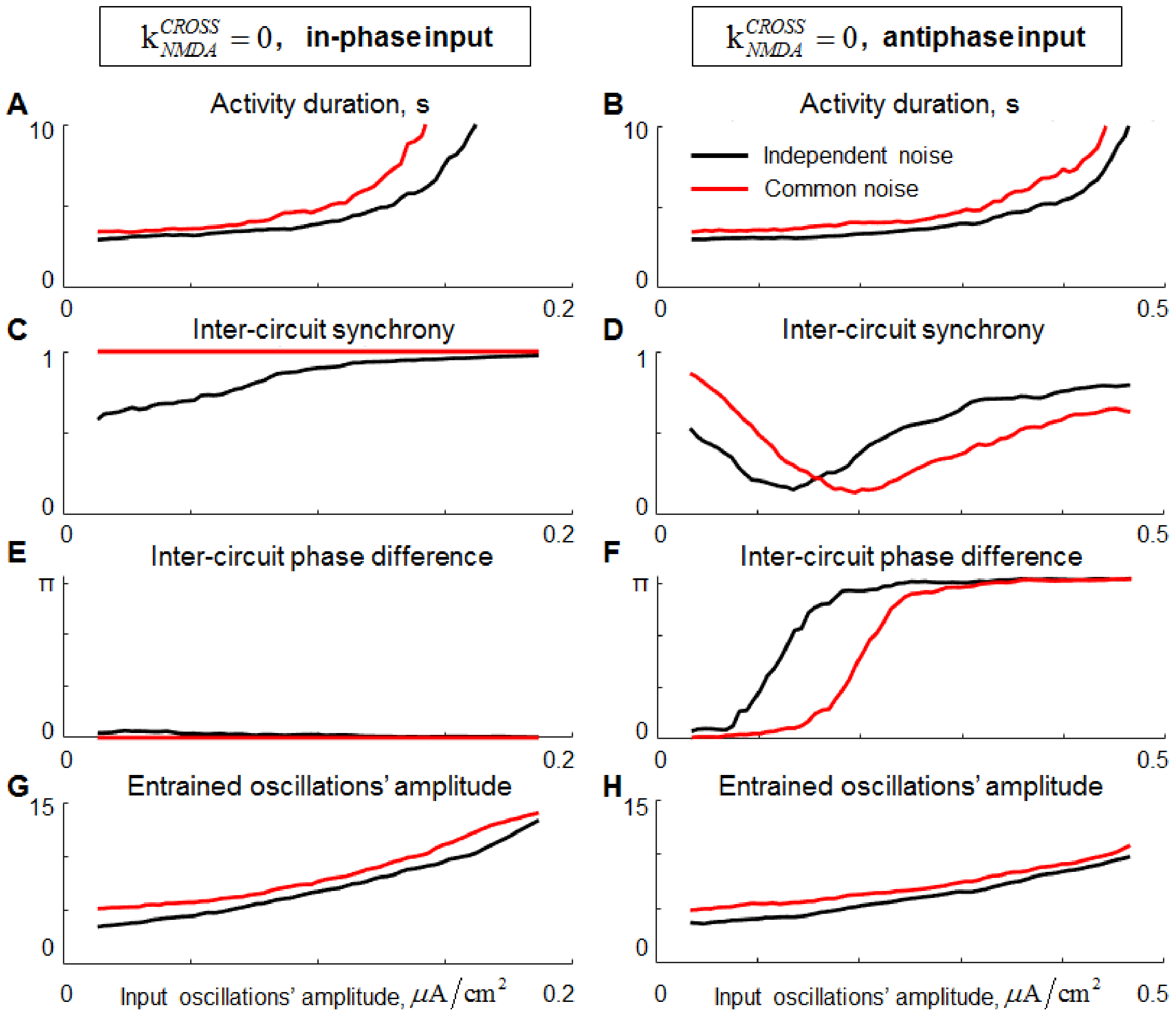

Figure S2. Behavior of the two-circuit model with fast inter-circuit connections under input oscillations of different amplitudes. Left column: the circuits receive oscillations in the same phase, right column – in opposite phases. Black curves: the circuits receive independent noise signals; red curves – a common noise signal. A, B – post-stimulus activity duration, averaged over the two circuits. The activity of a circuit is considered to be terminated when the firing rate of the excitatory population, smoothed with 100-ms time window, falls below 3 Hz. C, D – inter-circuit synchrony measured as the phase-locking value between the excitatory firing rate signals. E, F – Mean phase difference between the excitatory firing rate signals. G, H – Mean amplitude of the entrained oscillations of the excitatory firing rate signal. For calculation of C – H, the firing rate dynamics of the excitatory population of each circuit was bandpass-filtered (30 – 50 Hz), after which the Hilbert transform was applied to extract the amplitude and phase of the filtered signal at each time moment.

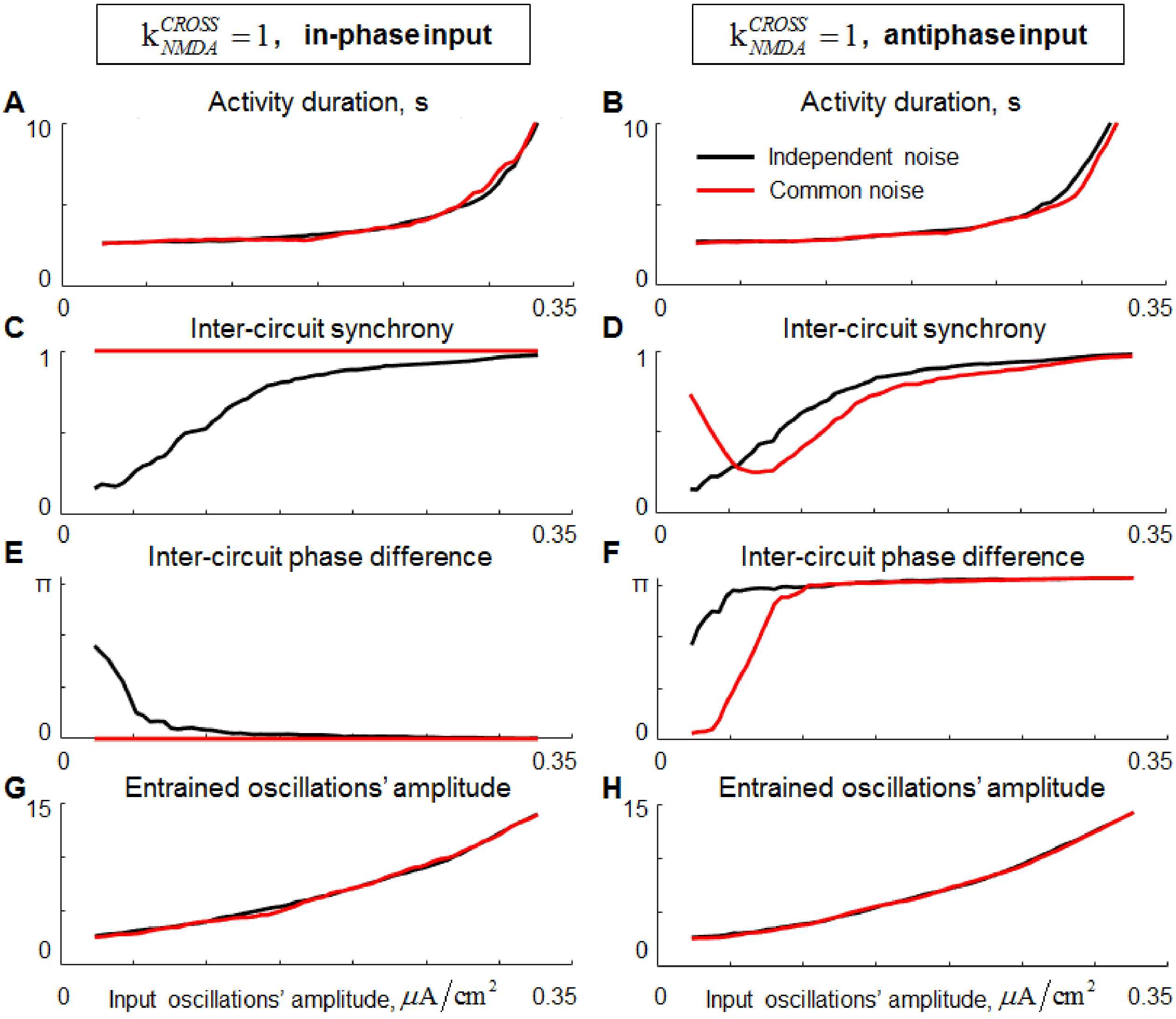

Figure S3. Behavior of the two-circuit model with slow inter-circuit connections under input oscillations of different amplitudes. The legend is the same as in Fig. S2.